\documentclass[12pt,titlepage,letterpaper]{article}
\usepackage[dvips]{graphicx}
\usepackage{amsfonts}
\title{A Local Ensemble Kalman Filter for Atmospheric Data Assimilation}
\date{Submitted to Monthly Weather Review \\ Revised \today}

\author{Edward Ott,\thanks{{\it Corresponding author:} Institute for
Research in Electronics and Applied Physics,
                                    University of Maryland,
                                    College Park, MD, 20742-2431,
                                    \mbox{E-mail:eo4@umail.umd.edu}}
        Brian R. Hunt,
        Istvan Szunyogh,
        Aleksey V. Zimin,\\
        Eric J. Kostelich,
        Matteo Corazza,
        Eugenia Kalnay,
        D.J. Patil,\\
        and James A. Yorke\\
        {\small \it University of Maryland
                    College Park, Maryland, USA;}
}
\begin{document}
\maketitle
\begin{abstract}
In this paper, we introduce a new, local formulation of the ensemble Kalman Filter
approach for atmospheric data assimilation. Our scheme is based on the hypothesis
that, when the Earth's surface is divided up into local regions of moderate size,
vectors of the forecast uncertainties in such regions tend to lie in a subspace of
much lower dimension than that of the full atmospheric state vector of such a
region. Ensemble Kalman Filters, in general, assume that the analysis resulting from
the data assimilation lies in the same subspace as the expected forecast error.
Under our hypothesis the dimension of this subspace is low. This implies that
operations only on  relatively low dimensional matrices are required. Thus, the data
analysis is done locally in a manner allowing massively parallel
computation to be exploited. The local analyses are then used to construct global
states for advancement to the next forecast time. The method, its potential
advantages, properties, and implementation requirements are illustrated by numerical
experiments on the Lorenz-96 model.  It is found that accurate analysis can be
achieved at a cost which is very modest compared to that of a full global ensemble
Kalman Filter.
\end{abstract}

\section{Introduction}
The purpose of this paper is to develop and test a new atmospheric
data assimilation scheme, which we call the Local Ensemble Kalman
Filter method.
Atmospheric \emph{data assimilation} (\emph{analysis}) is the process
through which an estimate of the atmospheric state is obtained by using
observed data and a dynamical model of the atmosphere (e.g., Daley 1991;
Kalnay 2002). These estimates, called \emph{analyses}, can then be used as
initial conditions in operational numerical weather predictions. In
addition, diagnostic studies of atmospheric dynamics and climate
are also often based on analyses instead of raw observed data.

The analysis at a given time instant is an approximation to a maximum likelihood
estimate of the atmospheric state in which a short-term forecast, usually referred
to as the \emph{background} or \emph{first guess field}, is used as a \emph{prior}
estimate of the atmospheric state (Lorenc 1986). Then the observations are
assimilated into the background field by a statistical interpolation. This
interpolation is performed based on the assumptions \emph{(i)} that the
uncertainties in the background field and the observations are unbiased and normally
distributed, \emph{(ii)} that there are no cross correlations between background and
observations errors, and \emph{(iii)} that the covariance between different
components of the background (formally the \emph{the background covariance matrix})
and the covariances between uncertainties in the noisy observations (formally the
\emph{observational error covariance matrix}) are known. In reality, however, the
background error covariance matrix cannot be directly computed. The implementation
of a data assimilation system, therefore, requires the development of statistical
models that can provide an estimate of the background error covariance matrix. The
quality of a data assimilation system is primarily determined by the accuracy of
this estimate.

In the case of linear dynamics the mathematically consistent technique to define an
adaptive background covariance matrix is the Kalman Filter (Kalman 1960; Kalman and
Bucy 1961) which utilizes the dynamical equations to evolve the most probable state
and the error covariance matrix in time. In the case of linear systems with unbiased
normally distributed errors the Kalman Filter provides estimates of the system state
that are optimal in the mean square sense.  The method has also been adapted to
nonlinear systems, but, in this case, optimality no longer applies.  Although the
Kalman Filter approach has been successfully implemented for a wide range of
applications and has been considered for atmospheric data assimilation for a long
while (Jones 1965; Petersen 1973; Ghil et al. 1981, Dee et al., 1985), the
computational cost involved does not allow for an operational implementation in the
foreseeable future (see Daley 1991 for details).

The currently most popular approach to reduce the cost of the Kalman Filter is to
use a relatively small (10-100 member) ensemble of background forecasts to estimate
the background error covariances (e.g. Evensen 1994; Houtekamer and Mitchell 1998,
2001; Bishop et al. 2001; Hamill et al. 2001; Whitaker and Hamill 2002; Keppenne and
Rienecker 2002). In ensemble-based data assimilation schemes the ensemble of
background forecasts is generated by using initial conditions distributed according
to the result of the previous analysis. [In this paper we will not consider the
issue of model error.  How to appropriately incorporate model error in an ensemble
Kalman filter, especially when accounting for the fact that such errors are
temporally correlated, is still an open question.]

The ensemble-based approach has the appeal of providing initial
ensemble perturbations that are consistent with the analysis scheme. This
is important because currently implemented operational techniques generate
initial ensemble perturbations without use of direct information about the
analysis errors (Toth and Kalnay 1997; Molteni et al. 1996). These
techniques are obviously suboptimal considering the goal of ensemble
forecasting, which is to simulate the effect of the analysis uncertainties
on the ensuing forecasts.

The main difference between the existing ensemble-based schemes is in the generation
of the analysis ensemble. One family of schemes is based on \emph{perturbed
observations} (Evensen and van Leeuwen 1996; Houtekamer and Mitchell 1998, 2001;
Hamill and Snyder 2000, 2001, Keppenne and Rienecker 2002).
 In this
approach, the analysis ensemble is obtained by assimilating a different set of
observations to each member of the background ensemble. The different sets of
observations are created by adding random noise to the real observations, where the
random noise component is generated according to the observational error covariance
matrix. The main weakness of this approach is that the ensemble size must be large
in order to accurately represent the probability distribution of the background
errors. Thus a relatively large forecast ensemble has to be evolved in time,
limiting the efficiency of the approach. Recent papers have discussed how
 the required size of the ensemble can be reduced (Houtekamer and Mitchell 1998 and 2001; Hamill et al. 2001),
 e.g., by filtering the long distance covariances from the
background field.

The other family of schemes, the Kalman square-root filters, use a different
approach to reduce the size of the ensemble. These techniques do the analysis only
once, to obtain the mean analysis. Then the analysis ensemble perturbations (to the
mean analysis) are generated by linearly transforming the background ensemble
perturbations to a set of vectors that can be used to represent the \emph{analysis
error covariance matrix}.  Thus, the analysis is confined to the subspace of the
ensemble.  This type of Kalman square-root strategy is feasible because the analysis
error covariance matrix can be computed explicitly from the background and
observational error covariance matrices. Since there is an infinite set of analysis
perturbations that can be used to represent the analysis error covariance matrix,
many different schemes can be derived following this approach (Tippett et al.,
2002). Existing examples of the square root filter approach are the Ensemble
Transform Kalman Filter (Bishop et al. 2001), the Ensemble Adjustment Filter
(Anderson 2001), and the Ensemble Square-root Filter (Whitaker and Hamill 2001).

The scheme we propose is a Kalman square-root filter \footnote{The basic algorithm
was first described in the paper Ott, E., B. H. Hunt, I. Szunyogh, M. Corazza, E.
Kalnay, D. J. Patil, J. A. Yorke, A. V. Zimin, and E. Kostelich, 2002: ``Exploiting
local low dimensionality of the atmospheric dynamics for efficient Kalman
filtering'' (http://arxiv.org/abs/physics/0203058).}. The most important difference
between our scheme and the other Kalman square-root filters is that  our analysis is
done locally in model space.  In a sense, our paper is related to
previous work that attempted to construct a simplified Kalman Filter by explicitly
taking into account the dominant unstable directions of the state space (Kalnay and
Toth 1994; Fisher 1998).

Our scheme is based on the construction of local regions about each grid point. The
basic idea is that we do the analysis at each grid point using the state variables
and observations in the local region centered at that point.  The effect is similar
to using a covariance localization filter (e.g., Hamill et al. 2001) whose
characteristic length is roughly the size of our local regions. An outline of the
scheme is as follows.
\begin{enumerate}
\item Advance the analysis ensemble of global atmospheric states to the next
analysis time, thus obtaining a new background ensemble of global atmospheric
states.

\item For each local region and each member of the background ensemble, form vectors
of the atmospheric state information in that local region. (Section 2) \item In each
local region, project the `local' vectors, obtained in step~2, onto a low
dimensional subspace that best represents the ensemble in that region. (Section 2)
\item Do the data assimilation in each of the local low dimensional subspaces,
obtaining analyses in each local region. (Section 3) \item Use the local analyses,
obtained in step~4, to form a new global analysis ensemble. (This is where the
square root filter comes in.) (Section 4)
 \item Go back
to step~1.
\end{enumerate}
These steps are summarized, along with a key of important symbols that we use, in
Figure \ref{fig:1_a} and its caption.

This method is potentially advantageous in that the individual local analyses
are done in low dimensional subspaces, so that matrix operations involve only
relatively low dimensional matrices. Furthermore, since the individual analyses
in different local regions do not interact, they can be done independently
in parallel.

In the following sections we describe and test our new approach to data
assimilation. Section~2 introduces the concept of local regions and explains how the
dimension of the local state vector can be further reduced. Section~3 explains the
analysis scheme for the local regions. In section~4, the local analyses are pieced
together to obtain the global analysis field and the ensemble of global analysis
perturbations. Section~5 illustrates our data assimilation scheme by an application
to a toy spatio-temporally chaotic model system introduced by Lorenz (1996).

\section{Local vectors and their covariance}

A model state of the atmosphere is given by a vector field
$\mathbf{x} (\mathbf{r},t)$ where $\mathbf{r}$ is two
dimensional and runs over discrete values $\mathbf{r}_{mn}$ (the grid in the
physical space used in the numerical computations). Typically, the two
components of $\mathbf{r}$ are the geographical longitude and latitude, and
$\mathbf{x}$ at a fixed $\mathbf{r}$ is a vector of all relevant physical
state variables of the model (e.g., wind velocity components, temperature,
surface pressure, humidity, etc., at all height levels included in the model).
Let $u$ denote the dimensionality of $\mathbf{x}(\mathbf{r},t)$ (at fixed
$\mathbf{r}$); e.g., when five independent state variables are defined at 28
vertical levels, $u=140$.

Data assimilation schemes generally treat $\mathbf{x} (\mathbf{r},t)$ as a random
variable with a time-dependent probability distribution. The characterization of
$\mathbf{x}$ is updated over time in two ways: (i) it is evolved according to the
model dynamics; and (ii) it is modified periodically to take into account recent
atmospheric observations.

We do our analysis locally in model space. In this section we introduce our local
coordinate system and the approximations we make to the local probability
distribution of $\mathbf{x} (\mathbf{r},t)$. Since all the analysis operations take
place at a fixed time $t$, we will suppress the $t$ dependence of all vectors and
matrices introduced henceforth.

Motivated by the work of Patil et al. (2001) we introduce at each point \emph{local
vectors} $\mathbf{x}_{mn}$ of the information $\mathbf{x}(\mathbf{r}_{m+m',n+n'},t)$
for $-l \leq m',n' \leq l$. That is, $\mathbf{x}_{mn}$ specifies the model
atmospheric state within a $(2l+1)$ by $(2l+1)$ patch of grid points centered at
$\mathbf{r}_{mn}$. (This particular shape of the local region was chosen to keep the
notations as simple as possible, but different (e.g., circular) shape regions and
localization in the vertical direction can also be considered.) The dimensionality
of $\mathbf{x}_{mn}$ is $(2l+1)^2u$. We represent the construction of local vectors
via a linear operator $\mathbf{M}_{mn}$,
\begin{equation} \label{eq:equation1}
\mathbf{x}_{mn}=\mathbf{M}_{mn} \mathbf{x}(\mathbf{r},t).
\end{equation}
We now consider local
vectors obtained from the model as forecasts, using initial conditions
distributed according to the result of the previous analysis, and we
denote these by $\mathbf{x}^b_{mn}$ (where the superscript $b$ stands for
``background'').
Let $F_{mn}(\mathbf{x}^b_{mn})$ be our approximation to the
probability density function for $\mathbf{x}^b_{mn}$ at the current
analysis time $t$.
A fundamental assumption is that this probability distribution can be
usefully approximated
as Gaussian,
\begin{equation} \label{eq:gaussian}
F_{mn} (\mathbf{x}^b_{mn}) \sim \exp \big[-\frac{1}{2}(\mathbf{x}^b_{mn}-
\bar{\mathbf{x}}^b_{mn})^T (\mathbf{P}^b_{mn})^{-1}(\mathbf{x}^b_{mn}-
\bar{\mathbf{x}}^b_{mn}) \big],
\end{equation}
where $\mathbf{P}^b_{mn}$ and $\bar{\mathbf{x}}^b_{mn}$ are the
\emph{local background error covariance matrix} and most probable state associated with
$F_{mn}(\mathbf{x}^b_{mn})$.
Graphically, the level set
\begin{equation}
F_{mn}(\mathbf{x}^b_{mn})=e^{-1/2}F_{mn}
(\bar{\mathbf{x}}^b_{mn})
\end{equation}
is an ellipsoid as illustrated in Figure~\ref{fig:1}. The equation of this
\emph{probability ellipsoid} is
\begin{equation}
(\mathbf{x}^b_{mn}-\bar{\mathbf{x}}^b_{mn})^T ({\mathbf{P}^b_{mn}})^{-1}
(\mathbf{x}^b_{mn}-\bar{\mathbf{x}}^b_{mn}) =1.
\end{equation}
We emphasize that the Gaussian form for the background probability distribution,
$F_{mn}(\mathbf{x}^b_{mn}$), is rigorously justifiable only for a linear system, but
not for a nonlinear system such as the atmosphere.

As explained subsequently, the rank of the $(2l+1)^2 u$ by $(2l+1)^2 u$ covariance
matrix $\mathbf{P}^b_{mn}$ for our approximate probability distribution function
$F_{mn}$ is much less than $(2l+1)^2u$. Let
\begin{equation}
k=\mbox{rank}(\mathbf{P}^b_{mn});
\end{equation}
($k=2$ in Figure~\ref{fig:1}). Thus $\mathbf{P}^b_{mn}$ has a $(2l+1)^2u-k$
dimensional null space $\bar{\mathbb{S}}_{mn}$ and the inverse
$(\mathbf{P}^b_{mn})^{-1}$ is defined for the component of the vectors
$(\mathbf{x}^b_{mn}-\bar{\mathbf{x}}^b_{mn})$ lying in the $k$ dimensional
 subspace $\mathbb{S}_{mn}$ orthogonal to $\bar{\mathbb{S}}_{mn}$

In the data assimilation procedure we describe in this paper, the
background error covariance matrix $\mathbf{P}^b_{mn}$ and the most
probable background state $\bar{\mathbf{x}}^b_{mn}$ are derived
from a $k'+1$ member ensemble of global state field vectors
$\big\{\mathbf{x}^{b(i)} (\mathbf{r},t) \big\}$, $i=1,2,\cdots,k'+1$;
$k' \geq k \geq 1$.
The most probable state is given by
\begin{equation} \label{eq:ensmean}
\bar{\mathbf{x}}_{mn}^b=\mathbf{M}_{mn}\big[{(k'+1)}^{-1} \sum_{i=1}^{k'+1}
\mathbf{x}^{b(i)}
(\mathbf{r},t)\big].
\end{equation}
To obtain the local background error covariance matrix $\mathbf{P}_{mn}^b$
that we use in our analysis, we first consider a matrix
$\mathbf{P}_{mn}^{b'}$ given by
\begin{equation} \label{eq:bcov}
\mathbf{P}_{mn}^{b'}={k'}^{-1} \sum_{i=1}^{k'+1}
\delta\mathbf{x}_{mn}^{b(i)} \big(\delta\mathbf{x}_{mn}^{b(i)}\big)^T,
\end{equation}
where the superscribed $T$ denotes transpose, and
\begin{equation} \label{eq:enspert}
\delta\mathbf{x}_{mn}^{b(i)}=\mathbf{M}_{mn}\mathbf{x}^{b(i)}(\mathbf{r},t)-
\bar{\mathbf{x}}^b_{mn}(\mathbf{r},t).
\end{equation}
It is also useful to introduce the notation
\begin{equation} \label{eq:xmatrix}
\mathbf{X}_{mn}^b=(k')^{-1/2}\big[\delta\mathbf{x}_{mn}^{b(1)} \mid
\delta\mathbf{x}_{mn}^{b(2)}\mid \cdots \mid
\delta\mathbf{x}_{mn}^{b(k'+1)}\big],
\end{equation}
in terms of which (\ref{eq:bcov}) can be rewritten,
\begin{equation} \label{eq:bcov2}
\mathbf{P}_{mn}^{b'}=\mathbf{X}_{mn}^b \mathbf{X}_{mn}^{bT}.
\end{equation}
We assume that forecast uncertainties in the mid-latitude extra-tropics tend to lie
in a low dimensional subset of the $(2l+1)^2u$ dimensional local vector space
\footnote{Preliminary results with an implementation of our data assimilation scheme
on the NCEP GFS supports this view.}. Thus we anticipate that we can approximate the
background error covariance matrix by one of much lower rank than $(2l+1)^2u$, and
this motivates our assumption that an ensemble of size of $k'+1$, where $k'+1$ is
substantially less than $(2l+1)^2u$, will be sufficient to yield a good approximate
representation of the background covariance matrix. Typically,
$\mathbf{P}^{b'}_{mn}$ has rank $k'$, i.e., it has $k'$ positive eigenvalues. Let
the eigenvalues of the matrix $\mathbf{P}^{b'}_{mn}$ be denoted by
$\lambda^{(j)}_{mn}$, where the labeling convention for the index $j$ is
\begin{equation} \label{eq:eigenlist}
\lambda^{(1)}_{mn} \ge \lambda^{(2)}_{mn} \ge \ldots \ge
\lambda^{(k)}_{mn} \ge \cdots \ge \lambda^{(k')}_{mn}.
\end{equation}
Since $\mathbf{P}^{b'}_{mn}$ is a symmetric matrix, it has $k'$ orthonormal
eigenvectors
\big\{$\mathbf{u}^{(j)}_{mn}$\big\} corresponding to the $k'$ eigenvalues
(\ref{eq:eigenlist}). Thus
\begin{equation} \label{eq:eq10}
\mathbf{P}^{b'}_{mn}=\sum^{k'}_{j=1}\lambda^{(j)}_{mn}\mathbf{u}^{(j)}_{mn}(\mathbf{u}^{(j)}_{mn})^T.
\end{equation}
Since the size of the ensemble is envisioned to be much less than the dimension of
$\mathbf{x}_{mn}^b$, $(k'+1) \ll (2l+1)^2u$, the computation of the eigenvalues and
eigenvectors of $\mathbf{P}_{mn}^{b'}$ is most effectively done in the basis of the
ensemble vectors. That is, we consider the eigenvalue problem for the $(k'+1) \times
(k'+1)$ matrix $\mathbf{X}_{mn}^{bT} \mathbf{X}_{mn}^b$, whose nonzero eigenvalues
are those of $\mathbf{P}_{mn}^{b'}$ [\ref{eq:eigenlist}] and whose corresponding
eigenvectors left-multiplied by $\mathbf{X}_{mn}^b$ are the $k'$ eigenvectors
$\mathbf{u}_{mn}^{(j)}$ of $\mathbf{P}_{mn}^{b'}$. We approximate
$\mathbf{P}^{b'}_{mn}$ by truncating the sum at $k \le k'$
\begin{equation} \label{eq:eq11}
\mathbf{P}^{b}_{mn}=\sum^k_{j=1} \lambda^{(j)}_{mn} \mathbf{u}^{(j)}_{mn}
\big(\mathbf{u}^{(j)}_{mn}\big)^T.
\end{equation}
In terms of $\mathbf{u}^{(j)}_{mn}$ and $\lambda^{(j)}_{mn}$,
the principal axes of the probability ellipsoid (Figure~\ref{fig:1}) are
given by
\begin{equation}
\sqrt{\lambda^{(j)}_{mn}}\mathbf{u}^{(j)}_{mn}.
\end{equation}
The basic justification for the approximation of the covariance by
$\mathbf{P}^b_{mn}$ is our supposition that for reasonably small values of $k$, the
error variance in all other directions is  much less than the variance,
\begin{equation} \label{eq:supposition}
\sum^{k}_{j=1}\lambda^{(j)}_{mn},
\end{equation}
in the directions \{$\mathbf{u}^{(j)}_{mn}$\}, $j=1,2,\ldots,k$. The truncated
covariance matrix  $\mathbf{P}^{b}_{mn}$ is determined not only by the dynamics of
the model but also by the choice of the components of
$\delta\mathbf{x}_{mn}^{b(i)}$. In order to meaningfully compare eigenvalues,
Equation~(\ref{eq:eigenlist}), the different components of
$\delta\mathbf{x}_{mn}^{b(i)}$ (e.g., wind and temperature) should be properly
scaled to ensure that, if the variance (\ref{eq:supposition}) approximates the full
variance, then the first $k$ eigendirections, \{$\mathbf{u}^{(j)}_{mn}$\},
$j=1,2,\ldots,k$, explain the important uncertainties in the background,
$\bar{\mathbf{x}}^b_{mn}$. For instance, the weights for the different variables can
chosen so that the Euclidean norm of the transformed vectors is equal to their
\emph{total energy norm} derived in Talagrand (1981). In what follows, we assume
that the vector components are already properly scaled. (We also note that if
$k=k'$, the comparison of eigenvalues is not used and thus such a consistent scaling
of the variables is not necessary.)

For the purpose of subsequent computation, we consider the coordinate
system for the $k$ dimensional space $\mathbb{S}_{mn}$ determined by
the basis vectors $\{\mathbf{u}_{mn}^{(j)}\}$.  We call this the
\emph{internal coordinate system} for $\mathbb{S}_{mn}$.
To change between the internal coordinates and those of the local space,
we introduce the $(2l+1)^2u$ by $k$ matrix,
\begin{equation} \label{eq:eq14}
\mathbf{Q}_{mn}=\big\{\mathbf{u}_{mn}^{(1)}{} \vert
\mathbf{u}_{mn}^{(2)} \vert \cdots \vert \mathbf{u}_{mn}^{(k)} \big\}.
\end{equation}
We denote the
projection of vectors into $\mathbb{S}_{mn}$ and the restriction of
matrices to $\mathbb{S}_{mn}$ by a superscribed circumflex (hat).
Thus
for a $(2l+1)^2u$ dimensional column vector $\mathbf{w}$, the vector
$\hat{\mathbf{w}}$ is a $k$ dimensional column vector given by
\begin{equation}
\hat{\mathbf{w}} = \mathbf{Q}^T_{mn} \mathbf{w}.
\end{equation}
Note that this operation consists of both projecting $\mathbf{w}$ into
$\mathbb{S}_{mn}$ and changing to the internal coordinate system.
Similarly, for a $(2l+1)^2u$ by $(2l+1)^2u$ matrix $\mathbf{M}$, the matrix
$\hat{\mathbf{M}}$ is $k$ by $k$ and given by
\begin{equation}
\hat{\mathbf{M}} = \mathbf{Q}^T_{mn}
\mathbf{M} \mathbf{Q}_{mn}.
\end{equation}
To go back to the original $(2l+1)^2u$ dimensional local vector space, note that
$\mathbf{Q}^T_{mn} \mathbf{Q}_{mn} = \mathbf{I}$ while $\mathbf{Q}_{mn}
\mathbf{Q}^T_{mn}$ represents projection on $\mathbb{S}_{mn}$, i.e., it has null
space $\bar{\mathbb{S}}_{mn}$ and acts as the identity on $\mathbb{S}_{mn}$. We may
write $\mathbf{w}$ as
\begin{equation} \label{eq:19}
\mathbf{w}=\mathbf{w}^{(\parallel)}+\mathbf{w}^{(\perp)},
\end{equation}
\begin{equation} \label{eq:20}
\mathbf{w}^{(\parallel)}=\mathbf{\Lambda}_{mn}^{(\parallel)}\mathbf{w}=
\mathbf{Q}_{mn}\hat{\mathbf{w}}, \qquad
\mathbf{w}^{(\perp)}=\mathbf{\Lambda}_{mn}^{(\perp)} \mathbf{w},
\end{equation}
where $\mathbf{w}^{(\parallel)}$ and $\mathbf{w}^{(\perp)}$ denote the
components of $\mathbf{w}$ in $\mathbb{S}_{mn}$ and $\bar{\mathbb{S}}_{mn}$,
respectively, and the projection operators
$\mathbf{\Lambda}_{mn}^{(\parallel)}$ and $\mathbf{\Lambda}_{mn}^{(\perp)}$ are
given by
\begin{equation} \label{eq:eq21}
\mathbf{\Lambda}_{mn}^{(\parallel)}=\mathbf{Q}_{mn}\mathbf{Q}_{mn}^T, \qquad
\mathbf{\Lambda}_{mn}^{(\perp)}=\mathbf{I}-\mathbf{Q}_{mn}\mathbf{Q}_{mn}^T.
\end{equation}
In addition, if $\mathbf{M}$ is symmetric with null space
$\bar{\mathbb{S}}_{mn}$,
\begin{equation}
\mathbf{M}=\mathbf{Q}_{mn}\hat{\mathbf{M}}\mathbf{Q}^T_{mn}.
\end{equation}
Note that $\hat{\mathbf{P}}^b_{mn}$ is diagonal,
\begin{equation} \label{eq:eq19}
\hat{\mathbf{P}}^b_{mn}=\mbox{diag} \big(\lambda_{mn}^{(1)},
\lambda_{mn}^{(2)}, ..., \lambda_{mn}^{(k)}\big),
\end{equation}
and thus it is trivial to invert.

\section{Data assimilation}
With Section~2 as background, we now consider the assimilation of observational
data to obtain a new specification of the probability distribution of the
local vector. In what follows, the notational convention of Ide et al. (1997)
is adopted whenever it is possible.

Let $\mathbf{x}^a_{mn}$ be the random variable at the current analysis time $t$
representing the local vector after knowledge of the observations and background
mean are taken into account. For simplicity, we assume that all observations
collected for the current analysis were taken at the same time $t$. Let
$\mathbf{y}^o_{mn}$ be the vector of current observations within the local region,
and assume that the errors in these observations are unbiased, are uncorrelated with
the background, and are normally distributed with covariance matrix
$\mathbf{R}_{mn}$. An ideal (i.e., noiseless) measurement is a function of the true
atmospheric state. Considering measurements within the local region $(m,n)$, we
denote this function $\mathbf{\mathcal{H}}_{mn}(\cdot)$. That is, if the true local
state is $\mathbf{x}^a_{mn}$, then the error in the observation is
$\mathbf{y}_{mn}^o- \mathbf{\mathcal{H}}_{mn}(\mathbf{x}_{mn}^a)$. Assuming that the
true state is near the mean background state $\bar{\mathbf{x}}_{mn}^b$, we
approximate $\mathbf{\mathcal{H}}_{mn}(\mathbf{x}_{mn}^a)$ by linearizing about
$\bar{\mathbf{x}}_{mn}^b$,
\begin{equation}
\mathbf{\mathcal{H}}_{mn}(\mathbf{x}_{mn}^a) \approx
\mathbf{\mathcal{H}}_{mn}(\bar{\mathbf{x}}_{mn}^b) + \mathbf{H}_{mn}\Delta
\mathbf{x}_{mn}^a,
\end{equation}
where
\begin{equation}
\Delta \mathbf{x}_{mn}^a = \mathbf{x}_{mn}^a - \bar{\mathbf{x}}_{mn}^b,
\end{equation}
and the matrix $\mathbf{H}_{mn}$ is the Jacobian matrix of partial derivatives
of $\mathbf{\mathcal{H}}_{mn}$ evaluated at $\bar{\mathbf{x}}_{mn}^b$.
(If there are $s$ scalar observations in the local $(2l+1)$ by $(2l+1)$ region
at analysis time $t$, then $\bar{\mathbf{y}}^o_{mn}$ is $s$ dimensional and
the rectangular matrix $\mathbf{H}_{mn}$ is $s$ by $(2l+1)^2u$). Then, since
we have assumed the background (pre-analysis) state $\mathbf{x}_{mn}^b$ to be
normally distributed, it will follow below that $\mathbf{x}_{mn}^a$ is also
normally distributed. Its distribution is determined by the most probable
state $\bar{\mathbf{x}}^a_{mn}$ and the associated covariance matrix
$\mathbf{P}^a_{mn}$. The data assimilation step determines
$\bar{\mathbf{x}}^a_{mn}$ (the \emph{local analysis}) and $\mathbf{P}^a_{mn}$
(the \emph{local analysis covariance matrix}).

Since our approximate background covariance matrix  $\mathbf{P}^b_{mn}$
has null space $\bar{\mathbb{S}}_{mn}$, we consider the analysis increment
component $\Delta\mathbf{x}^{a(\parallel)}_{mn} =
\Lambda_{mn}^{(\parallel)}
\big(\mathbf{x}^a_{mn}-\bar{\mathbf{x}}^b_{mn}\big)$ within the
$k$-dimensional subspace $\mathbb{S}_{mn}$, and do the data assimilation
in $\mathbb{S}_{mn}$. Thus the data assimilation is done by minimizing the
quadratic form,
\begin{eqnarray} \label{eq:costfunction}
J\big(\Delta\hat{\mathbf{x}}^a_{mn}\big) & = &
\big(\Delta\hat{\mathbf{x}}^a_{mn}\big)^T
\big(\hat{\mathbf{P}}^b_{mn}\big)^{-1}
\Delta\hat{\mathbf{x}}^a_{mn} \nonumber \\
 & + &
\big(\hat{\mathbf{H}}_{mn}
\Delta\hat{\mathbf{x}}^a_{mn}+
\mathbf{\mathcal{H}}_{mn}
(\bar{\mathbf{x}}^b_{mn})
-\mathbf{y}^o_{mn}\big)^T
\mathbf{R}_{mn}^{-1} \times \nonumber \\
& & \big(\hat{\mathbf{H}}_{mn}
\Delta\hat{\mathbf{x}}^a_{mn}+
\mathbf{\mathcal{H}}_{mn}
(\bar{\mathbf{x}}^b_{mn})
-\mathbf{y}^o_{mn}\big).
\end{eqnarray}
Here $\hat{\mathbf{H}}_{mn} = \mathbf{H}_{mn} \mathbf{Q}_{mn}$
maps $\mathbb{S}_{mn}$ to the observation space, using
the internal coordinate system for $\mathbb{S}_{mn}$ introduced in the
previous section, so that $\Delta\mathbf{x}^{a(\parallel)}_{mn} =
\mathbf{Q}_{mn}\Delta\hat{\mathbf{x}}^a_{mn}$.
The most probable value of $\Delta\hat{\mathbf{x}}^a_{mn}$,
\begin{eqnarray} \label{eq:analysis}
\Delta\hat{\bar{\mathbf{x}}}^a_{mn}  =
\hat{\mathbf{P}}^a_{mn} \hat{\mathbf{H}}^T_{mn}\mathbf{R}_{mn}^{-1}
\big(\mathbf{y}^o_{mn}-\mathbf{\mathcal{H}}_{mn}(\bar{\mathbf{x}}^b_{mn})\big),
\end{eqnarray}
is the minimizer of $J\big(\Delta\hat{\mathbf{x}}^a_{mn}\big)$,
where the analysis covariance matrix $\hat{\mathbf{P}}^a_{mn}$ is
the inverse of the matrix of second derivatives (Hessian) of
$J\big(\Delta\hat{\mathbf{x}}^a_{mn}\big)$ with respect to
$\Delta\hat{\mathbf{x}}^a_{mn}$,
\begin{equation} \label{eq:analysis2}
\hat{\mathbf{P}}^a_{mn} = \big[\big(\hat{\mathbf{P}}^b_{mn}\big)^{-1}+
\hat{\mathbf{H}}^T_{mn}\mathbf{R}_{mn}^{-1}\hat{\mathbf{H}}_{mn}
\big]^{-1}.
\end{equation}
For computational purposes, we prefer to use the alternate form,
\begin{equation} \label{eq:an2alt}
\hat{\mathbf{P}}^a_{mn} = \hat{\mathbf{P}}^b_{mn} \big[\mathbf{I} +
\hat{\mathbf{H}}^T_{mn}\mathbf{R}_{mn}^{-1}\hat{\mathbf{H}}_{mn}
\hat{\mathbf{P}}^b_{mn}\big]^{-1},
\end{equation}
both in place of (\ref{eq:analysis2}) and in computing
(\ref{eq:analysis}). A potential numerical advantage of (\ref{eq:an2alt}) over
(\ref{eq:analysis2}) is that (\ref{eq:analysis2}) involves the inverse
of $\hat{\mathbf{P}}^b_{mn}$, which may be problematic if
$\hat{\mathbf{P}}^b_{mn}$ has a small eigenvalue.

Another alternative is to compute (\ref{eq:analysis}) and
(\ref{eq:analysis2}) in  terms of the ``Kalman gain'' matrix
\begin{equation} \label{eq:kalgain}
\hat{\mathbf{K}}_{mn} = \hat{\mathbf{P}}^b_{mn}
\hat{\mathbf{H}}^T_{mn} \big(\hat{\mathbf{H}}_{mn}
\hat{\mathbf{P}}^b_{mn} \hat{\mathbf{H}}^T_{mn} +
\mathbf{R}_{mn}\big)^{-1}.
\end{equation}
Then it can be shown (e.g., Kalnay 2002, p. 171) that (\ref{eq:analysis}) and
(\ref{eq:analysis2})/(\ref{eq:an2alt}) are equivalent to
\begin{equation}
\Delta\hat{\bar{\mathbf{x}}}^a_{mn} = \hat{\mathbf{K}}_{mn}
\big(\mathbf{y}^o_{mn}-\mathbf{H}_{mn}\bar{\mathbf{x}}^b_{mn}\big),
\end{equation}
and
\begin{equation} \label{eq:kal3}
\hat{\mathbf{P}}^a_{mn} = \big(\mathbf{I} - \hat{\mathbf{K}}_{mn}
\hat{\mathbf{H}}_{mn}\big) \hat{\mathbf{P}}^b_{mn}.
\end{equation}
Again, the inverse of $\hat{\mathbf{P}}^b_{mn}$ is not required.

Though (\ref{eq:analysis}) and (\ref{eq:an2alt}) are mathematically
equivalent to (\ref{eq:kalgain})--(\ref{eq:kal3}), the former approach
may be significantly more efficient computationally for the following
reasons.  In both cases, one must invert an $s$ by $s$ matrix, where
$s$ is the number of local observations.  While these matrices are
considerably smaller than those involved in global data assimilation
schemes, they may still be quite large.  Generally the $s$ by $s$
matrix $\mathbf{R}_{mn}$ whose inverse is required in (\ref{eq:an2alt})
will be diagonal or close to diagonal, and thus less expensive to
invert than the matrix inverted in (\ref{eq:kalgain}).  (Furthermore,
in some cases one may be able to treat $\mathbf{R}_{mn}$ as
time-independent and avoid recomputing its inverse for each successive
analysis.)  The additional inverse required in (\ref{eq:an2alt}) is of
a $k$ by $k$ matrix, where $k \leq k'$ may be relatively small compared
to $s$ if the number of observations in the local region $(m,n)$ is large.

Finally, going back to the local space representation, we have
\begin{equation}
\bar{\mathbf{x}}^a_{mn}=\mathbf{Q}_{mn} \Delta\hat{\bar{\mathbf{x}}}^a_{mn} +
\bar{\mathbf{x}}^b_{mn}.
\end{equation}

\section{Updating the ensemble}
We now wish to use the analysis information, $\hat{\mathbf{P}}^a_{mn}$ and
$\bar{\mathbf{x}}^a_{mn}$, to obtain an ensemble of global analysis fields
$\big\{\mathbf{x}^{a(i)}(\mathbf{r},t) \big\}$; $i=1,2,\cdots,k'+1$. Once these
fields are determined, they can be used as initial conditions for the
atmospheric model. Integrating these global fields forward in time to the
next analysis time $t+\Delta t$, we obtain the background ensemble
$\big\{\mathbf{x}^{b(i)}(\mathbf{r},t+\Delta t)\big\}$. This
completes the loop, and, if the procedure is stable, it can be repeated
for as long as desired. Thus at each analysis time we are in possession
of a global initial condition that can be used for making forecasts of
the desired durations.

Our remaining task is to specify the ensemble of global analysis fields
$\big\{\mathbf{x}^{a(i)}(\mathbf{r},t)\big\}$ from our
analysis information, $\hat{\mathbf{P}}^a_{mn}$ and
$\bar{\mathbf{x}}^a_{mn}$.
Denote $(k'+1)$ local analysis vectors by
\begin{equation} \label{eq:eq4_1}
    \mathbf{x}^{a(i)}_{mn} = \bar{\mathbf{x}}^a_{mn}+
    \delta \mathbf{x}^{a(i)}_{mn}.
\end{equation}
Using (\ref{eq:19}) and (\ref{eq:20}) we write
\begin{equation} \label{eq:aeq4_1}
\delta \mathbf{x}_{mn}^{a(i)}=\delta \mathbf{x}_{mn}^{a(i)(\parallel)} +
\delta \mathbf{x}_{mn}^{a(i)(\perp)}=\mathbf{Q}_{mn} \delta
\hat{\mathbf{x}}_{mn}^{a(i)}+\delta \mathbf{x}_{mn}^{a(i)(\perp)}.
\end{equation}
In addition, we let
\begin{equation} \label{eq:perpcomp}
\delta \mathbf{x}_{mn}^{a(i)(\perp)} = \delta
\mathbf{x}_{mn}^{b(i)(\perp)}= \Lambda_{mn}^{(\perp)} \delta
\mathbf{x}^{b(i)}_{mn},
\end{equation}
because our analysis uses the observations only to reduce the variance in the space
$\mathbb{S}_{mn}$, leaving the variance in $\bar{\mathbb{S}}_{mn}$ unchanged. (We
note, however, that by our construction of $\bar{\mathbb{S}}_{mn}$ in section~2, the
total variance in $\bar{\mathbb{S}}_{mn}$ is expected to be small compared to that
in $\mathbb{S}_{mn}$. Also, in the case $k=k'$ all members of the analysis
perturbation ensemble will lie in $\mathbb{S}_{mn}$, so that projection onto
$\mathbb{S}_{mn}$ is superfluous, and $\delta \mathbf{x}_{mn}^{a(i)(\perp)}$ in
(\ref{eq:aeq4_1}) and the term $\Lambda_{mn}^{(\perp)} \delta
\mathbf{x}^{b(i)}_{mn}$ in  (\ref{eq:latest}) (below) may be omitted.) Combining
(\ref{eq:20}) and (\ref{eq:eq4_1})-(\ref{eq:perpcomp}), we have
\begin{equation} \label{eq:latest}
\mathbf{x}_{mn}^{a(i)}=\bar{\mathbf{x}}_{mn}^a + \mathbf{Q}_{mn} \delta
\hat{\mathbf{x}}^{a(i)}_{mn} + \Lambda_{mn}^{(\perp)} \delta \mathbf{x}^{b(i)}_{mn}.
\end{equation}
We require that
\begin{equation} \label{eq:38new_label}
\sum_{i=1}^{k'+1} \delta \mathbf{x}_{mn}^{a(i)}=\mathbf{0},
\end{equation}
which, by virtue of (\ref{eq:perpcomp}), and (from (\ref{eq:ensmean}) and
(\ref{eq:38new_label}))
\begin{equation} \label{eq:37}
\sum_{i=1}^{k'+1} \delta \mathbf{x}_{mn}^{b(i)}=\mathbf{0},
\end{equation}
is equivalent to
\begin{equation}
\sum_{i=1}^{k'+1} \delta \mathbf{x}_{mn}^{a(i)(\parallel)}=
\mathbf{Q}_{mn} \sum_{i=1}^{k'+1} \delta
\hat{\mathbf{x}}_{mn}^{a(i)}=\mathbf{0}.
\end{equation}
Thus we require that
\begin{equation} \label{eq:eq4_2}
    \sum_{i=1}^{k'+1}\delta\hat{\mathbf{x}}^{a(i)}_{mn}=\mathbf{0}.
\end{equation}
In addition,
$\hat{\mathbf{P}}^a_{mn}$ is given by
\begin{equation} \label{eq:eq4_3}
    \hat{\mathbf{P}}^a_{mn}={k'}^{-1}\sum_{i=1}^{k'+1}\delta\hat{\mathbf{x}}^{a(i)}_{mn}
    \big(\delta\hat{\mathbf{x}}^{a(i)}_{mn}\big)^T.
\end{equation}
Hence the local analysis state $\bar{\mathbf{x}}^{a}_{mn}$
(determined in Section~3) is the mean over the local analysis ensemble
$\big\{\mathbf{x}^{a(i)}_{mn}\big\}$,
and, by (\ref{eq:eq4_3}), $\big\{\delta
\hat{\mathbf{x}}^{a(i)}_{mn}\big\}$ gives a representation of the local
analysis error covariance matrix. We now turn to the task of determining
the analysis perturbations $\big\{\delta
\hat{\mathbf{x}}^{a(i)}_{mn}\big\}$. Once these are known
$\big\{\mathbf{x}^{a(i)}_{mn}\big\}$ is determined from (\ref{eq:latest}).

\subsection{Determining the ensemble of local analysis perturbations}
There are many choices for $\big\{\delta\hat{\mathbf{x}}^{a(i)}_{mn}\big\}$ that
satisfy (\ref{eq:eq4_2}) and (\ref{eq:eq4_3}), and in this section we will describe
possible methods for computing a set of solutions to these equations.  (See also
Tippett et al. (2002) for different approaches to this problem in the global
setting.)  In a given forecasting scenario, one could compare the accuracy and speed
of these methods in order to choose among them.  There are two main criteria we have
in mind in formulating these methods.

First, the method for computing
$\big\{\delta\hat{\mathbf{x}}^{a(i)}_{mn}\big\}$ should be numerically
stable and efficient.
Second, since we wish to specify global
fields that we think of as being similar to physical fields, we desire that
these fields be slowly varying in $m$ and $n$.  That is, if
$\hat{\mathbf{P}}^a_{mn}$ is slowly varying, we do not want to introduce
any artificial rapid variations in the individual
$\delta\hat{\mathbf{x}}^{a(i)}_{mn}$
through our method of constructing a solution of (\ref{eq:eq4_2}) and
(\ref{eq:eq4_3}). For this purpose we regard the background vectors as
physical states, and hence slowly varying in $m$ and $n$.  (This is
reasonable since the background ensemble is obtained from evolution
of the atmospheric model from time $t-\Delta t$ to time $t$.)

Thus we are motivated to express the analysis ensemble vectors
$\delta\hat{\mathbf{x}}^{a(i)}_{mn}$ as formally linearly related to the
background ensemble vectors. We consider two possible methods for doing
this. In the first method, we relate $\delta\hat{\mathbf{x}}^{a(i)}_{mn}$
to the background vector with the same label $i$,
\begin{equation} \label{eq:eq4_4}
    \delta\hat{\mathbf{x}}^{a(i)}_{mn}=\mathbf{Z}_{mn}
        \delta\hat{\mathbf{x}}^{b(i)}_{mn},
\end{equation}
where
\begin{equation} \label{eq:new2}
\delta\hat{\mathbf{x}}^{b(i)}_{mn}=
\mathbf{Q}^T_{mn}\delta\mathbf{x}^{b(i)}_{mn}.
\end{equation}
(Note that the apparent linear relation between the background and analysis
perturbations in (\ref{eq:eq4_4}) is only formal, since our solution for
$\mathbf{Z}_{mn}$ will depend on the background perturbations.) In the second
method, we formally express $\delta\hat{\mathbf{x}}^{a(i)}_{mn}$ as a linear
combination of the vectors, $\delta\hat{\mathbf{x}}^{b(1)}_{mn}$,
$\delta\hat{\mathbf{x}}^{b(2)}_{mn}$, $\cdots$,
$\delta\hat{\mathbf{x}}^{b(k'+1)}_{mn}$,
\begin{equation} \label{eq:rightmult}
\hat{\mathbf{X}}^a_{mn} = \hat{\mathbf{X}}^b_{mn} \mathbf{Y}_{mn}.
\end{equation}
where
\begin{equation} \label{eq:Xdef}
\hat{\mathbf{X}}^{a,b}_{mn} = \big(k'\big)^{-1/2}
\big\{\delta\hat{\mathbf{x}}^{a,b(1)}_{mn} \vert
\delta\hat{\mathbf{x}}^{a,b(2)}_{mn} \vert \cdots \vert
\delta\hat{\mathbf{x}}^{a,b(k'+1)}_{mn} \big\}.
\end{equation}
Using (\ref{eq:Xdef}) the analysis and the background covariance matrices
can be expressed as
\begin{equation} \label{eq:new}
\hat{\mathbf{P}}^{a,b}=\hat{\mathbf{X}}^{a,b}_{mn}
\hat{\mathbf{X}}^{a,bT}_{mn}.
\end{equation}
The $k \times k$ matrix $\mathbf{Z}_{mn}$
or the $(k'+1) \times (k'+1)$ matrix $\mathbf{Y}_{mn}$ can be thought of as a
generalized `rescaling' of the original background fields.  This
`rescaling' can be viewed as being similar to the techniques employed
in the breeding method (Toth and Kalnay, 1993) and in the Ensemble
Transform Kalman Filter approach (Bishop et al., 2001; Wang and Bishop,
2002).  If $\mathbf{Z}_{mn}$ or $\mathbf{Y}_{mn}$
vary slowly with $m$ and $n$, then by (\ref{eq:eq4_4}) and
(\ref{eq:rightmult}) so will $\delta\hat{\mathbf{x}}^{a(i)}_{mn}$.

Considering (\ref{eq:eq4_4}), we see that (\ref{eq:eq4_2}) is automatically
satisfied because, by (\ref{eq:37}) and (\ref{eq:new2}), the background
perturbations $\delta\hat{\mathbf{x}}^{b(i)}_{mn}$ sum to zero,
\begin{equation} \label{eq:v}
\hat{\mathbf{X}}^b_{mn}\mathbf{v}=\mathbf{0},
\end{equation}
where $\mathbf{v}$ is a column vector of $(k'+1)$ ones.
The analysis perturbations given by (\ref{eq:eq4_4}) will satisfy
(\ref{eq:eq4_3}) [equivalently (\ref{eq:new})] if, and only if,
\begin{equation} \label{eq:Z}
\hat{\mathbf{P}}^a_{mn}=\mathbf{Z}_{mn} \hat{\mathbf{P}}^b_{mn}
\mathbf{Z}^T_{mn}.
\end{equation}

Considering (\ref{eq:rightmult}), we see that (\ref{eq:new}) yields the
following equation for $\mathbf{Y}_{mn}$
\begin{equation} \label{eq:Y}
\hat{\mathbf{P}}^a_{mn}=\hat{\mathbf{X}}^b_{mn} \mathbf{Y}_{mn} \mathbf{Y}^T_{mn}
\hat{\mathbf{X}}^{bT}_{mn}.
\end{equation}
Unlike (\ref{eq:eq4_4}) for $\mathbf{Z}_{mn}$, (\ref{eq:rightmult}) does not
imply automatic satisfaction of (\ref{eq:eq4_2}). We note that (\ref{eq:eq4_2})
can be written as
\begin{equation}
\hat{\mathbf{X}}^a_{mn}\mathbf{v}=\mathbf{0}.
\end{equation}
Thus, in addition to ({\ref{eq:Y}), we demand that $\mathbf{Y}_{mn}$ must also
satisfy
\begin{equation} \label{eq:Y2}
\hat{\mathbf{X}}^b_{mn}\mathbf{Y}_{mn} \mathbf{v} = \mathbf{0}.
\end{equation}

Equation (\ref{eq:Z}) has infinitely many solutions for $\mathbf{Z}_{mn}$.
Similarly, equations (\ref{eq:Y}) and (\ref{eq:Y2}) have infinitely
many solutions for $\mathbf{Y}_{mn}$. In order for the results to vary slowly
from one grid point to the next, it is important that we use an algorithm
for computing a particular solution that depends continuously on
$\hat{\mathbf{P}}^a_{mn}$ and $\hat{\mathbf{P}}^b_{mn}$.

\subsection{Solutions of Equation (\ref{eq:Z})}
\subsubsection{Solution 1}
One solution $\mathbf{Z}_{mn}$ is
\begin{equation} \label{eq:Solution1}
\mathbf{Z}_{mn}=\big(\hat{\mathbf{P}}^a_{mn}\big)^{1/2}
\big(\hat{\mathbf{P}}^b_{mn}\big)^{-1/2},
\end{equation}
where in (\ref{eq:Solution1}), by the notation $\mathbf{M}^{1/2}$, we mean
the \emph{unique} positive symmetric square root of the positive symmetric
matrix $\mathbf{M}$. In terms of the eigenvectors and eigenvalues of
$\mathbf{M}$, the positive symmetric square root is
\begin{equation}
\mathbf{M}^{1/2}=\sum_{j=1}^k \sqrt{\nu^{(j)}} \mathbf{m}^{(j)}
\big(\mathbf{m}^{(j)}\big)^T,
\end{equation}
where
\begin{equation}
\mathbf{M} \mathbf{m}^{(j)} = \nu^{(j)} \mathbf{m}^{(j)}.
\end{equation}
Recall that $\hat{\mathbf{P}}^b_{mn}$ is diagonal, so that its inverse
square root in (\ref{eq:Solution1}) is easily computed.

\subsubsection{Solution 2}
Pre- and post-multiplying (\ref{eq:Z}) by
$\big(\hat{\mathbf{P}}^b_{mn}\big)^{1/2}$ and taking
$\mathbf{Z}_{mn}$ to be symmetric,
\begin{equation} \label{eq:prepost}
\Big[\big(\hat{\mathbf{P}}^b_{mn} \big)^{1/2} \mathbf{Z}_{mn}
\big(\hat{\mathbf{P}}^b_{mn} \big)^{1/2} \big]^2=
\big(\hat{\mathbf{P}}^b_{mn} \big)^{1/2} \hat{\mathbf{P}}^a_{mn}
\big(\hat{\mathbf{P}}^b_{mn} \big)^{1/2}.
\end{equation}
Taking the positive symmetric square root of (\ref{eq:prepost}),
we obtain a second possible solution of (\ref{eq:Z}),
\begin{equation} \label{eq:Solution2}
\mathbf{Z}_{mn}=\big(\hat{\mathbf{P}}^b_{mn} \big)^{-1/2}
\Big[
\big(\hat{\mathbf{P}}^b_{mn} \big)^{1/2} \hat{\mathbf{P}}^a_{mn}
\big(\hat{\mathbf{P}}^b_{mn} \big)^{1/2}
\Big]^{1/2} \big(\hat{\mathbf{P}}^b_{mn} \big)^{-1/2}.
\end{equation}
In contrast to solution~1 (given by (\ref{eq:Solution1})) and solutions~3
(given below), this solution yields a $\mathbf{Z}_{mn}$ that is symmetric,
$\mathbf{Z}_{mn}=\mathbf{Z}^T_{mn}$.

\subsubsection{Family of solutions}
We can create a family of solutions for $\mathbf{Z}_{mn}$ by introducing
an arbitrary positive symmetric matrix $\mathbf{D}_{mn}$ and by pre- and
post-multiplying (\ref{eq:Z}) by $\mathbf{D}_{mn}^{-1/2}$. This yields
\begin{equation} \label{eq:gsola}
\tilde{\mathbf{P}}_{mn}^{a}=\tilde{\mathbf{Z}}_{mn}
\tilde{\mathbf{P}}_{mn}^{b}
\tilde{\mathbf{Z}}_{mn}^T,
\end{equation}
where
\begin{equation} \label{eq:gsolb}
\tilde{\mathbf{Z}}_{mn}=\mathbf{D}_{mn}^{-1/2} \mathbf{Z}_{mn}
\mathbf{D}_{mn}^{1/2},
\end{equation}
\begin{equation} \label{eq:gsolc}
\tilde{\mathbf{P}}_{mn}^{a,b}=\mathbf{D}_{mn}^{-1/2}
\hat{\mathbf{P}}_{mn}^{a,b} \mathbf{D}_{mn}^{-1/2}.
\end{equation}
Applying solution~2 to (\ref{eq:gsola}) we obtain (\ref{eq:Solution2}) with
$\mathbf{Z}_{mn}$ and $\hat{\mathbf{P}}_{mn}^{a,b}$ replaced by
$\tilde{\mathbf{Z}}_{mn}$ and $\tilde{\mathbf{P}}_{mn}^{a,b}$, Then,
applying (\ref{eq:gsolb}) and (\ref{eq:gsolc}), we find that the unique
solution to (\ref{eq:Z}) such that $\mathbf{D}_{mn}^{-1/2} \mathbf{Z}_{mn}
\mathbf{D}_{mn}^{1/2}$ is symmetric is
\begin{equation} \label{eq:GS1}
\mathbf{Z}_{mn}=
\mathbf{D}_{mn}^{1/2}\big(\tilde{\mathbf{P}}_{mn}^b\big)^{-1/2}
\Big[\big(\tilde{\mathbf{P}}_{mn}^b\big)^{-1/2} \tilde{\mathbf{P}}_{mn}^a
\big(\tilde{\mathbf{P}}_{mn}^b\big)^{-1/2}\Big]^{1/2}
\big(\tilde{\mathbf{P}}_{mn}^b\big)^{-1/2} \mathbf{D}_{mn}^{-1/2}.
\end{equation}
Thus for any choice of $\mathbf{D}_{mn}$ we obtain a solution
$\mathbf{Z}_{mn}$ of (\ref{eq:Z}), and this is the unique solution for
$\mathbf{Z}_{mn}$ subject to the added condition that
$\mathbf{D}_{mn}^{-1/2} \mathbf{Z}_{mn} \mathbf{D}_{mn}^{1/2}$ is symmetric.

Another way to generate a family of solutions is to replace
(\ref{eq:Solution1}) by
\begin{equation}
    \mathbf{Z}_{mn}=\sqrt{\hat{\mathbf{P}}^a_{mn}}
    \sqrt{\big(\hat{\mathbf{P}}^b_{mn}\big)^{-1}}^T,
\end{equation}
where for a positive definite symmetric matrix $\mathbf{M}$,
we mean by $\sqrt{\mathbf{M}}$ any matrix for which
$\sqrt{\mathbf{M}} \sqrt{\mathbf{M}}^T = \mathbf{M}$.  Note that this
equation does not uniquely determine $\sqrt{\mathbf{M}}$, and that
given any solution $\sqrt{\mathbf{M}} = \mathbf{W}$, the most general
solution is $\sqrt{\mathbf{M}} = \mathbf{W}\mathbf{O}$ where
$\mathbf{O}$ is any orthogonal matrix.  In particular, the positive symmetric
square root (which we denote $\mathbf{M}^{1/2}$) is a specific choice
for $\sqrt{\mathbf{M}}$, and, in general, $\sqrt{\mathbf{M}}=\mathbf{M}^{1/2}
\mathbf{O}$. Furthermore, by considering
all possible matrices $\sqrt{\hat{\mathbf{P}}^a_{mn}}$ we obtain
all possible solutions $\mathbf{Z}_{mn}$ of
(\ref{eq:Z}). Thus we can write a general solution of (\ref{eq:Z}) as
\begin{equation} \label{eq:GS2}
\mathbf{Z}_{mn}=\big(\hat{\mathbf{P}}_{mn}^a\big)^{1/2} \mathbf{O}_{mn}
\big(\hat{\mathbf{P}}_{mn}^{b} \big)^{-1/2},
\end{equation}
where $\mathbf{O}_{mn}$ is an arbitrary orthogonal matrix. (Note that
$\mathbf{O}_{mn}$ can be a function of $\hat{\mathbf{P}}^a_{mn}$ and
$\hat{\mathbf{P}}^b_{mn}$.) For further discussion see Appendix A.

The family of solutions of (\ref{eq:Z}) generated by (\ref{eq:GS1}) with different
$\mathbf{D}_{mn}$ is smaller than the family given by (\ref{eq:GS2}) with different
$\mathbf{O}_{mn}$. In particular, the family (\ref{eq:GS2}), being the most general
solution of (\ref{eq:Z}), must contain the family corresponding to (\ref{eq:GS1}).
To see that the latter family is indeed smaller than the former family, consider the
special case, $\hat{\mathbf{P}}_{mn}^a=\hat{\mathbf{P}}_{mn}^b$. For
$\hat{\mathbf{P}}_{mn}^a=\hat{\mathbf{P}}_{mn}^b$, (\ref{eq:GS1}) always gives
$\mathbf{Z}_{mn}=\mathbf{I}$, while (\ref{eq:GS2}) gives
\begin{equation} \label{eq:GS3}
\mathbf{Z}_{mn}=\big(\hat{\mathbf{P}}_{mn}^a\big)^{-1/2} \mathbf{O}_{mn}^{(o)}
\big(\hat{\mathbf{P}}_{mn}^{a} \big)^{1/2},
\end{equation}
which is never $\mathbf{I}$ unless the orthogonal matrix
$\mathbf{O}^{(o)}_{mn}$ is $\mathbf{I}$. (Here $\mathbf{O}^{(o)}_{mn}$
denotes $\mathbf{O}_{mn}$ evaluated at
$\hat{\mathbf{P}}_{mn}^a=\hat{\mathbf{P}}_{mn}^b$.) Based on our
treatment in section~4.3, we believe that the smaller family, given by
(\ref{eq:GS1}) with different $\mathbf{D}_{mn}$, gives results for
$\hat{\mathbf{X}}_{mn}^a$ that are more likely to be useful for our purposes.

\subsubsection{Solution~3}
Subsequently, special interest will attach to the choices $\mathbf{D}_{mn}=
\hat{\mathbf{P}}_{mn}^b$ and $\mathbf{D}_{mn}= \hat{\mathbf{P}}_{mn}^a$
in (\ref{eq:GS1}). Although these two choices yield results from
(\ref{eq:gsolc}) and (\ref{eq:GS1}) that appear to be of quite different
form,
the two results for $\mathbf{Z}_{mn}$ are in fact the same. We call this
solution for $\mathbf{Z}_{mn}$ solution~3. To see that these two
$\mathbf{D}_{mn}$ choices yield the same $\mathbf{Z}_{mn}$, we note that
(\ref{eq:Z}) can be put in the form,
\begin{equation}
\big(\hat{\mathbf{P}}_{mn}^a\big)^{1/2} \Big[\big(
\hat{\mathbf{P}}_{mn}^a\big)^{-1/2}
\mathbf{Z}_{mn} \big(\hat{\mathbf{P}}_{mn}^{a}\big)^{1/2} \Big]^{-1}
\big(\hat{\mathbf{P}}_{mn}^{a}\big)^{1/2}
=
\big(\hat{\mathbf{P}}_{mn}^b\big)^{1/2}
\Big[\big(\hat{\mathbf{P}}_{mn}^b\big)^{-1/2}
\mathbf{Z}_{mn} \big(\hat{\mathbf{P}}_{mn}^b\big)^{1/2} \Big]^T
\big(\hat{\mathbf{P}}_{mn}^{b}\big)^{1/2}.
\end{equation}
Thus symmetry of $\big(\hat{\mathbf{P}}_{mn}^a\big)^{-1/2}
\mathbf{Z}_{mn} \big(\hat{\mathbf{P}}_{mn}^{a}\big)^{1/2}$ (required by
the choice $\mathbf{D}_{mn}=\mathbf{P}_{mn}^a$ in (\ref{eq:GS1})) implies
symmetry of
$\big(\hat{\mathbf{P}}_{mn}^b\big)^{-1/2} \mathbf{Z}_{mn}
\big(\hat{\mathbf{P}}_{mn}^b\big)^{1/2}$ (i.e.,
$\mathbf{D}_{mn}=\hat{\mathbf{P}}_{mn}^b$ in (\ref{eq:GS1})) and vice
versa. Hence, the two choices for $\mathbf{D}_{mn}$ necessarily yield the
same $\mathbf{Z}_{mn}$. Explicitly, setting $\mathbf{D}_{mn}=
\hat{\mathbf{P}}_{mn}^b$ in (\ref{eq:GS1}) we can write solution~3 as
\begin{equation} \label{eq:Solution3}
\mathbf{Z}_{mn}=\big(\hat{\mathbf{P}}_{mn}^b \big)^{1/2} \Big[
\big(\hat{\mathbf{P}}_{mn}^b\big)^{-1/2}\hat{\mathbf{P}}_{mn}^a
\big(\hat{\mathbf{P}}_{mn}^b\big)^{-1/2}\Big]^{1/2}
\big(\hat{\mathbf{P}}_{mn}^b\big)^{-1/2}.
\end{equation}
As discussed subsequently, alternate formulations exist for which solution 3 does
not require inverting $\hat{\mathbf{P}}^b_{mn}$ (see Equations (77), (79), and
(80)). This may be advantageous if $\hat{\mathbf{P}}^b_{mn}$ has small eigenvalues.

\subsection{`Optimal' choices for $\mathbf{Z}_{mn}$}
Since we think of the background ensemble members as physical fields, it is
reasonable to seek to choose the analysis ensemble perturbations
$\delta \hat{\mathbf{x}}^{a(i)}_{mn}$ in such a way as to minimize their
difference with the background,
\begin{equation} \label{eq:quad1}
\mathcal{F}(\delta \hat{\mathbf{x}}^{a(i)}_{mn})=
\sum_{i=1}^{k'+1} \big\| \delta \hat{\mathbf{x}}^{a(i)}_{mn}-
\delta \hat{\mathbf{x}}^{b(i)}_{mn} \big\|^2=
\sum_{i=1}^{k'+1}
\big[\delta \hat{\mathbf{x}}^{a(i)}_{mn}-
\delta \hat{\mathbf{x}}^{b(i)}_{mn} \big]^T
\big[\delta \hat{\mathbf{x}}^{a(i)}_{mn}-
\delta \hat{\mathbf{x}}^{b(i)}_{mn} \big],
\end{equation}
subject to the requirement that
(\ref{eq:eq4_3}) be satisfied. Thus, introducing a $k \times k$ matrix
$\mathbf{B}_{mn}$ of Lagrange multipliers, we form the following quantity,
\begin{equation} \label{eq:Lagrange}
\mathcal{L}= \sum_{i=1}^{k'+1}
\big[ \delta \hat{\mathbf{x}}^{a(i)}_{mn}-
\delta \hat{\mathbf{x}}^{b(i)}_{mn} \big]^T
\big[ \delta \hat{\mathbf{x}}^{a(i)}_{mn}-
\delta \hat{\mathbf{x}}^{b(i)}_{mn} \big]-
\sum_{p,q=1}^k (\mathbf{B}_{mn})_{p,q} \Big[ \big(\hat{\mathbf{P}}^a_{mn}
\big)_{p,q} - \frac{1}{k'} \sum_{i=1}^{k'+1} \big( \delta
\hat{\mathbf{x}}^{a(i)}_{mn} \big)_p \big( \delta
\hat{\mathbf{x}}^{a(i)}_{mn} \big)_q \Big]
\end{equation}
which we minimize with respect to $\delta \hat{\mathbf{x}}^{a(i)}_{mn}$ and
$\mathbf{B}_{mn}$. Forming the first and second derivatives of $\mathcal{L}$
with respect to $\delta \hat{\mathbf{x}}^{a(i)}_{mn}$, we have
\begin{equation} \label{eq:firstd}
\frac{1}{2} \frac{\partial \mathcal{L}}{\partial \delta
\hat{\mathbf{x}}^{a(i)}_{mn}}=
\mathbf{Z}^{-1}_{mn} \delta \hat{\mathbf{x}}^{a(i)}_{mn}- \delta
\hat{\mathbf{x}}^{b(i)}_{mn},
\end{equation}
\begin{equation} \label{eq:secondd}
\frac{1}{2} \frac{\partial^2 \mathcal{L}}{\partial \delta
\hat{\mathbf{x}}^{a(i)}_{mn}
\partial \delta \hat{\mathbf{x}}^{a(i)}_{mn}}= \mathbf{Z}^{-1}_{mn},
\end{equation}
where we have defined $\mathbf{Z}^{-1}_{mn}$ as
\begin{equation} \label{eq:zsym}
\mathbf{Z}^{-1}_{mn} = \mathbf{I} + \frac{1}{2k'}
\big(\mathbf{B}_{mn}+\mathbf{B}^T_{mn}\big).
\end{equation}
Since $\mathcal{L}$ is stationary, (\ref{eq:firstd}) implies
(\ref{eq:eq4_4}), and the derivative with respect to $\mathbf{B}_{mn}$
returns (\ref{eq:eq4_3}). Since $\mathcal{L}$ is minimum,
(\ref{eq:secondd}) implies that
$\mathbf{Z}_{mn}$ is positive, while (\ref{eq:zsym}) gives $\mathbf{Z}_{mn}=
\mathbf{Z}^T_{mn}$. Thus the solution that minimizes $\mathcal{F}
(\hat{\mathbf{x}}^{a(i)}_{mn})$ is obtained from the \emph{unique} symmetric
positive solution for $\mathbf{Z}_{mn}$. This is given by solution~2
(\ref{eq:Solution2}).

It is also of interest to consider different metrics for the distance between
the analysis ensemble $\big\{ \delta \hat{\mathbf{x}}^{a(i)}_{mn} \big\}$
and the background ensemble
$\big\{ \delta \hat{\mathbf{x}}^{b(i)}_{mn} \big\}$. Thus we minimize
the quadratic form,
\begin{equation} \label{eq:quad2}
\mathcal{F}_D(\delta \hat{\mathbf{x}}^{a(i)}_{mn})=
\sum_{i=1}^{k'+1} \big\| \delta \hat{\mathbf{x}}^{a(i)}_{mn}-
\delta \hat{\mathbf{x}}^{b(i)}_{mn} \big\|^2_D=
\sum_{i=1}^{k'+1}
\big[\delta \hat{\mathbf{x}}^{a(i)}_{mn}-
\delta \hat{\mathbf{x}}^{b(i)}_{mn} \big]^T
\mathbf{D}^{-1}_{mn}
\big[\delta \hat{\mathbf{x}}^{a(i)}_{mn}-
\delta \hat{\mathbf{x}}^{b(i)}_{mn} \big],
\end{equation}
where the positive symmetric matrix $\mathbf{D}_{mn}$ specifies the metric.
(The quadratic form $\mathcal{F}(\delta \hat{\mathbf{x}}^{a(i)}_{mn})$
is the special case of $\mathcal{F}_D(\delta \hat{\mathbf{x}}^{a(i)}_{mn})$
when the metric is defined by
the identity matrix, $\mathbf{D}_{mn}=\mathbf{I}$). The introduction of the
metric matrix $\mathbf{D}_{mn}$ is equivalent to making the change of
variables,
$\tilde{\mathbf{X}}^{a,b}_{mn}=(\mathbf{D}_{mn})^{-1/2}
\hat{\mathbf{X}}^{a,b}_{mn}$.
Inserting this change of variables in (\ref{eq:Solution2}), we obtain
(\ref{eq:GS1}).

Solution~3, namely $\mathbf{D}_{mn}$ equal to $\hat{\mathbf{P}}_{mn}^b$ or
$\hat{\mathbf{P}}_{mn}^a$, appears to be favorable in that it provides a
natural intuitive normalizations for the distance. We thus conjecture that
solutions~3 may yield better performance than solutions~1~and~2.

\subsection{Solution of (\ref{eq:Y}) and (\ref{eq:Y2})} \label{sec:sol_of}
Another way of solving for the analysis fields is to use the `Potter method'
(e.g., Biermann 1977). To see how this solution is obtained, let
\begin{equation} \label{eq:potter1}
\mathbf{A}_{mn}=\mathbf{Y}_{mn}\mathbf{Y}^T_{mn}
\end{equation}
so that (\ref{eq:Y}) becomes
\begin{equation} \label{eq:potter2}
\hat{\mathbf{P}}^a_{mn}= \hat{\mathbf{X}}^b_{mn} \mathbf{A}_{mn}
\hat{\mathbf{X}}^{bT}_{mn}
\end{equation}
Because $\hat{\mathbf{P}}^a_{mn}$ is $k \times k$ and $\mathbf{A}_{mn}$ is
$(k'+1) \times (k'+1)$, there is a lot of freedom in choosing
$\mathbf{A}_{mn}$. It seems reasonable that, if the analysis covariance and
the background covariance are the same (i.e., $\hat{\mathbf{P}}^a_{mn}=
\hat{\mathbf{P}}^b_{mn})$, then the ensemble analysis perturbations should be
set equal to the ensemble background perturbations:
\begin{equation} \label{eq:potter3}
\mathbf{Y}_{mn}=\mathbf{I} \qquad \textrm{if} \qquad
\hat{\mathbf{P}}^a_{mn}=\hat{\mathbf{P}}^b_{mn}.
\end{equation}
A solution for $\mathbf{A}_{mn}$ consistent with
(\ref{eq:potter1})-(\ref{eq:potter3}) is
\begin{equation} \label{eq:potter4}
\mathbf{A}_{mn}=\mathbf{I}+\hat{\mathbf{X}}^{bT}_{mn} \big( \hat{\mathbf{P}}^b_{mn}
\big)^{-1} \big[\hat{\mathbf{P}}^a_{mn}-\hat{\mathbf{P}}^b_{mn} \big] \big(
\hat{\mathbf{P}}^b_{mn} \big)^{-1} \hat{\mathbf{X}}^{b}_{mn}.
\end{equation}
This solution for $\mathbf{A}_{mn}$ is symmetric and can also be shown to be
positive definite. Equation (\ref{eq:potter4}) yields
$\mathbf{A}_{mn}=\mathbf{I}$ if
$\hat{\mathbf{P}}^a_{mn}=\hat{\mathbf{P}}^b_{mn}$, as required by
(\ref{eq:potter1}) and (\ref{eq:potter3}), and satisfaction of
(\ref{eq:potter2}) by (\ref{eq:potter4}) can be verified by direct substitution
and making use of $\hat{\mathbf{P}}^b_{mn}=\hat{\mathbf{X}}^b_{mn}
\hat{\mathbf{X}}^{bT}_{mn}$. From (\ref{eq:potter1}) we have
$\mathbf{Y}_{mn}= \sqrt{\mathbf{A}_{mn}}$, and, if the positive symmetric
square root is chosen, then (\ref{eq:potter3}) is satisfied. Thus we have
as a possible solution
\begin{equation} \label{eq:Y3}
\mathbf{Y}_{mn}=(\mathbf{A}_{mn})^{1/2}.
\end{equation}
It remains to show that (\ref{eq:potter4}) and (\ref{eq:Y3}) also satisfies
(\ref{eq:Y2}).
By (\ref{eq:potter4}) and (\ref{eq:v}) we have $\mathbf{A}_{mn} \mathbf{v}
= \mathbf{v}$; i.e., $\mathbf{v}$ is an eigenvector of $\mathbf{A}_{mn}$ with
eigenvalue one. Since the positive square root is employed in (\ref{eq:Y3})
$\mathbf{v}$ is also an eigenvector of $\mathbf{Y}_{mn}$ with eigenvalue one.
Hence $\mathbf{X}^b_{mn} \mathbf{Y}_{mn} \mathbf{v} = \mathbf{X}^b_{mn}
\mathbf{v}$, which is identically zero by (\ref{eq:v}), thus satisfying
(\ref{eq:Y2}).

Potter's expression for $\mathbf{A}_{mn}$ is obtained by using (\ref{eq:kalgain})
and (\ref{eq:kal3}) in (\ref{eq:potter4}),
\begin{equation} \label{eq:Aold1}
\mathbf{A}_{mn}= \mathbf{I}-\hat{\mathbf{X}}_{mn}^{bT}
\hat{\mathbf{H}}_{mn}^T \big[\hat{\mathbf{H}}_{mn} \hat{\mathbf{P}}_{mn}^b
\hat{\mathbf{H}}_{mn}^T + \mathbf{R}_{mn} \big]^{-1} \hat{\mathbf{H}}_{mn}
\hat{\mathbf{X}}_{mn}^{b}.
\end{equation}
For (\ref{eq:Y3}) and (\ref{eq:Aold1}) the square root is taken of a
$k'+1$ by $k'+1$ matrix, but the inverse is of an $s$ by $s$ matrix, where
$s$ is the dimension of the local observation space. An equivalent way to
write (\ref{eq:Aold1}) in our setting is
\begin{equation}
\mathbf{A}_{mn} = \mathbf{I}-\hat{\mathbf{X}}_{mn}^{bT}\hat{\mathbf{V}}_{mn}
\hat{\mathbf{H}}_{mn}^T \mathbf{R}_{mn}^{-1}  \hat{\mathbf{H}}_{mn}
\hat{\mathbf{X}}_{mn}^{b},
\end{equation}
where
\begin{equation}
\hat{\mathbf{V}}_{mn}=\big[\mathbf{I}+\hat{\mathbf{H}}_{mn}^T \mathbf{R}_{mn}^{-1}
\hat{\mathbf{H}}_{mn} \hat{\mathbf{P}}_{mn}^b \big]^{-1}.
\end{equation}
Now aside from $\mathbf{R}_{mn}$, we need only invert a $k$ by $k$ matrix.
As previously discussed, although $\mathbf{R}_{mn}$ is $s$ by $s$, its inverse
is easily computed even when $s$ is much larger than $k$.

We now ask whether each solution $\mathbf{Z}_{mn}$  of (\ref{eq:Z}) has a
corresponding $\mathbf{Y}_{mn}$ such that $\mathbf{Z}_{mn} \hat{\mathbf{X}}_{mn}^b$
and $\hat{\mathbf{X}}_{mn}^b \mathbf{Y}_{mn}$ yield the same result for
$\hat{\mathbf{X}}_{mn}^a$. To see that they do, we note that the matrix
$\hat{\mathbf{X}}_{mn}^b$ (which consists of $k$ rows and $k'+1$ columns) has a
(nonunique) right inverse $\big(\hat{\mathbf{X}}_{mn}^b\big)^{-1}$ such that
$\hat{\mathbf{X}}_{mn}^b \big(\hat{\mathbf{X}}_{mn}^b\big)^{-1}= \mathbf{I}_k$,
where
\begin{equation} \label{eq:b-1}
\big(\hat{\mathbf{X}}_{mn}^b\big)^{-1} = \hat{\mathbf{X}}_{mn}^{bT}
\big(\hat{\mathbf{X}}_{mn}^b \hat{\mathbf{X}}_{mn}^{bT}\big)^{-1} +
\mathbf{E}_{mn}=
\hat{\mathbf{X}}^{bT}_{mn}\big(\hat{\mathbf{P}}^b_{mn}\big)^{-1}+
\mathbf{E}_{mn},
\end{equation}
and $\mathbf{E}_{mn}$ is any $k\times(k'+1)$ matrix for which
$\hat{\mathbf{X}}_{mn}^b \mathbf{E}_{mn}=\mathbf{0}_{mn}$. Thus, from
$\hat{\mathbf{X}}_{mn}^a=\mathbf{Z}_{mn}\hat{\mathbf{X}}_{mn}^b$, we have
\begin{equation} \label{eq:b'}
\hat{\mathbf{X}}_{mn}^a = \hat{\mathbf{X}}_{mn}^b
\big(\hat{\mathbf{X}}_{mn}^b\big)^{-1}
\mathbf{Z}_{mn} \hat{\mathbf{X}}_{mn}^b.
\end{equation}
From the definition of $\mathbf{Y}_{mn}$, $\hat{\mathbf{X}}_{mn}^a=
\hat{\mathbf{X}}_{mn}^b \mathbf{Y}_{mn}$, we see that (\ref{eq:b'}) and
(\ref{eq:b-1}) yield
\begin{equation} \label{eq:c'}
\mathbf{Y}_{mn} = \hat{\mathbf{X}}_{mn}^{bT}
\big(\hat{\mathbf{P}}_{mn}^b\big)^{-1}
\mathbf{Z}_{mn}\hat{\mathbf{X}}_{mn}^b+\mathbf{G}_{mn},
\end{equation}
where $\mathbf{G}_{mn}$ is any $(k'+1)$ by $(k'+1)$ matrix satisfying
$\hat{\mathbf{X}}_{mn}^b \mathbf{G}_{mn} = \mathbf{0}$. Since we desire
that $\mathbf{Y}_{mn}=\mathbf{I}_{k'+1}$, when
$\mathbf{Z}_{mn}=\mathbf{I}_{k}$,
a possible choice for $\mathbf{G}_{mn}$ is
\begin{equation} \label{eq:d'}
\mathbf{G}_{mn} = \mathbf{I}_{k'+1} - \hat{\mathbf{X}}_{mn}^{bT}
\big(\hat{\mathbf{P}}_{mn}^b\big)^{-1} \hat{\mathbf{X}}_{mn}^b.
\end{equation}
(We note that $\mathbf{G}_{mn}$ given by (\ref{eq:d'}) is a projection
operator, $(\mathbf{G}_{mn})^p=\mathbf{G}_{mn}$
for any integer exponent $p$.) Thus from (\ref{eq:c'}) and (\ref{eq:d'}),
a $\mathbf{Y}_{mn}$ corresponding to any solution $\mathbf{Z}_{mn}$ (e.g.,
solution~1,~2~or~3) is
\begin{equation} \label{eq:e'}
\mathbf{Y}_{mn}=\hat{\mathbf{X}}_{mn}^{bT}
\big(\hat{\mathbf{P}}_{mn}^b\big)^{-1}
(\mathbf{Z}_{mn}-\mathbf{I}_k) \hat{\mathbf{X}}_{mn}^b + \mathbf{I}_{k'+1}.
\end{equation}
Using (\ref{eq:e'}), (\ref{eq:Z}), and (\ref{eq:v}) it can be verified that
$\mathbf{Y}_{mn}\mathbf{Y}_{mn}^T=\mathbf{A}_{mn}$ with $\mathbf{A}_{mn}$
given by (\ref{eq:potter4}).
Thus $\mathbf{Y}_{mn}\mathbf{Y}_{mn}^T$
is the \emph{same} $(k'+1)\times(k'+1)$ matrix for all solutions
$\mathbf{Z}_{mn}$ (e.g., solutions 1,2, and 3). The general solution of
$\mathbf{Y}_{mn}\mathbf{Y}_{mn}^T=\mathbf{A}_{mn}$ is
\begin{equation} \label{eq:f'}
\mathbf{Y}_{mn}=\sqrt{\mathbf{A}_{mn}}=(\mathbf{A}_{mn})^{1/2} \mathbf{O}_{mn},
\end{equation}
where $\mathbf{O}_{mn}$ is an arbitrary orthogonal matrix. However, to ensure
that (\ref{eq:Y2}) is satisfied we also require that $\mathbf{O}_{mn}
\mathbf{v}= \pm \mathbf{v}$ (where $\mathbf{v}$ is a column vector of
$(k'+1)$ ones); i.e., that $\mathbf{v}$ is an eigenvector of
$\mathbf{O}_{mn}$ with eigenvalue $\pm 1$.. For example, $\mathbf{O}_{mn}$ can be any rotation
about $\mathbf{v}$. Thus there is still a large family of allowed orthogonal
matrices $\mathbf{O}_{mn}$. (Note that $\mathbf{O}_{mn}$ can depend on
$\hat{\mathbf{P}}_{mn}^a$ and $\hat{\mathbf{P}}_{mn}^b$, and that for
(\ref{eq:potter3}) to be satisfied, $\mathbf{O}_{mn}$ must be $\mathbf{I}$
whenever $\hat{\mathbf{P}}_{mn}^a=\hat{\mathbf{P}}_{mn}^b$.)
Hence we can think of the various solutions
for $\mathbf{Y}_{mn}$ either as being generated by (\ref{eq:GS1}) and
(\ref{eq:e'}) with different choices for the metric matrix
$\mathbf{D}_{mn}$, or as being generated by (\ref{eq:potter4}) and
(\ref{eq:f'}) with different choices for the orthogonal matrix
$\mathbf{O}_{mn}$.

Note that since $\big(\hat{\mathbf{P}}^b_{mn} \big)^{-1} \mathbf{Z}_{mn}$ is
symmetric for solution~3 (e.g., see (\ref{eq:Solution3})), the resulting
$\mathbf{Y}_{mn}$ from (\ref{eq:e'}) is symmetric and must therefore coincide
with (\ref{eq:Y3}).
That is, $\mathbf{Z}_{mn}\hat{\mathbf{X}}_{mn}^b$
with $\mathbf{Z}_{mn}$ given by (\ref{eq:Solution3}) and
$\hat{\mathbf{X}}_{mn}^b \mathbf{Y}_{mn}$ with $\mathbf{Y}_{mn}$ given by
(\ref{eq:potter4}) and (\ref{eq:Y3}) both yield the same result
for $\hat{\mathbf{X}}_{mn}^a$.
Also, in Appendix~B we show that
$\mathbf{Y}_{mn}$ as given by (\ref{eq:e'}) can be used to directly
obtain the analysis $\delta \mathbf{x}_{mn}^{a(i)}$ (note the absence of
the superscribed circumflex on $\delta \mathbf{x}_{mn}^{a(i)}$).

\subsection{Construction of the global fields}
Regardless of which of these solution methods for $\{\delta
\hat{\mathbf{x}}^{a(i)}_{mn}\}$ is chosen, by use of (\ref{eq:latest})  we now have
$(k'+1)$ local analyses $\mathbf{x}^{a(i)}_{mn}$ at each point $\mathbf{r}_{mn}$,
and it now remains to construct an ensemble of global fields
$\big\{\mathbf{x}^{a(i)}(\mathbf{r},t)\big\}$ that can be propagated forward in time
to the next analysis time. There are various ways of doing this. The simplest method
is to take the state of the global vector, $\mathbf{x}^{a(i)}$, at the point
$\mathbf{r}_{mn}$ directly from the local vector, $\mathbf{x}^{a(i)}_{mn}$, for the
local regions centered at $\mathbf{r}_{mn}$.  This approach uses only the analysis
results at the center point of each local region to form the global analysis
vectors. Another method (used in our numerical example of section~5) takes into
account atmospheric states at the point $\mathbf{r}_{mn}$ obtained from all the
local vectors $\mathbf{x}^{a(i)}_{m-m',n-n'}$
 $(|m'|\le l, |n'|\le l)$ that include the point $\mathbf{r}_{mn}$.
In particular, these states at $\mathbf{r}_{mn}$ are averaged to obtain
$\mathbf{x}^{a(i)}(\mathbf{r},t)$. In forming the average we weight the different
local regions with weights depending on $(m',n')$ such that the weights decrease
away from $\mathbf{r}_{mn}$. The motivation for this is that, if we obtain from
$\mathbf{x}_{mn}^{a(i)}$ state estimates at points in the local region $mn$, then
the estimates may be expected to be less accurate for points toward the edges of the
local region. Note that such averaging to obtain
$\mathbf{x}^{a(i)}(\mathbf{r}_{mn},t)$ has the effect of gradually decreasing the
influence of observations that are further from the point $\mathbf{r}_{mn}$ at which
$\mathbf{x}^{a(i)}(\mathbf{r}_{mn},t)$ is being estimated. In order to illustrate
this, consider the case where the weights are equal for $ |m'| \le l'$, $ |n'| \le
l'$, where $l' < l$, and zero for $l \ge |m'| >l'$, $l \ge |n'| >l'$ (this is what
we do in Section 5). That is, we only average over states obtained from local
regions whose centers are within an inner $(2l'+1) \times (2l'+1)$ square contained
in the $(2l+1) \times (2l+1)$ local region $(m,n)$, and we give those inner states
equal weights. See Figure \ref{fig:2} for an illustration of the case $l=5$, $l'=2$.
For the example in Figure \ref{fig:2}, an observation at a point within the $7
\times 7$ inner square would be contained within all the 25 local regions centered
at points within the inner $5 \times 5$ square, but observations outside the inner
$7 \times 7$ square are contained in fewer of the local regions, and observations
outside the $21 \times 21$ square centered at point $mn$ are in none of the 25 local
regions. Note also that in the case $l'=0$, we use only the analysis at the center
point of the local region (no averaging).  We do not believe that there are any
universal best values for the weights used to form the average; in a particular
scheme, the parameter $l'$ can be varied to test different weightings, or more
general weights can be considered.

\subsection{Variance inflation}
In past work on ensemble Kalman filters (Anderson and Anderson 1999; Whitaker and
Hamill 2002) it was found that inflating the covariance ($\mathbf{P}^a$ or
$\mathbf{P}^b$) by a constant factor on each analysis step, leads to more stable and
improved analyses. One rationale for doing this is to compensate for the effect of
finite sample size, which can be shown to, on average, underestimate the covariance.
In addition, in Section~5 and Appendix~C we will investigate the usefulness of
enhancing the probability of error in directions that formally show only very small
error probability (i.e., eigendirections corresponding to small eigenvalues of the
covariance matrices). Following such a modification of $\hat{\mathbf{P}}^a_{mn}$ or
$\hat{\mathbf{P}}^b_{mn}$, for consistency, we also make modifications to the
ensemble perturbations $\delta\hat{\mathbf{x}}^{a(i)}_{mn}$ or
$\delta\hat{\mathbf{x}}^{b(i)}_{mn}$ so as to preserve the relationship
(\ref{eq:new}). (Again, similar to the discussion in Section~4.2.3, the choice of
these modifications is not unique.)

In our numerical experiments in section~5 we will consider two methods
of variance inflation. One method, which we refer to as \emph{regular
variance inflation}, multiplies all background perturbations $\delta
\hat{\mathbf{x}}^{b(i)}_{mn}$ by a constant $(1+\delta)$. This corresponds to
multiplying $\hat{\mathbf{P}}^b_{mn}$ by $(1+\delta)^2$. This method has been
previously used by Anderson and Anderson (1999) and by Whitaker and Hamill
(2002). In
addition to this method, in Appendix~C we introduce a second variance
inflation method, which, as our results of section~5 indicate, may yield
superior performance. We refer to this method as \emph{enhanced variance
inflation}.

\section{Numerical experiments}
\subsection{Lorenz-96 model}
In this section we will test the skill of the proposed local ensemble Kalman Filter
scheme by carrying out Observing System Simulation Experiments (OSSE's) on the
Lorenz-96 (L96) model (Lorenz 1996; Lorenz and Emanuel 1998),
\begin{equation} \label{eq:lorenz}
\frac{d x_m}{dt}=(x_{m+1}-x_{m-2})x_{m-1}-x_m+F.
\end{equation}
Here, $m=1,\cdots,M$, where $x_{-1}=x_{M-1}$, $x_0=x_{M}$, and $x_{M+1}=x_1$. This
model mimics the time evolution of an unspecified scalar meteorological quantity,
$x$, at $M$  equidistant grid points along a latitude circle. We solve
(\ref{eq:lorenz}) with a fourth-order Runge-Kutta time integration scheme with a
time step of 0.05 non-dimensional unit (which may be thought of as nominally
equivalent to 6-h in real world time assuming that the characteristic time scale of
dissipation in the atmosphere is 5-days; see Lorenz 1996 for details). We emphasize
that this toy model, (\ref{eq:lorenz}), is very different from a full atmospheric
model, and that it can, at best, only indicate possible trends and illustrate
possible behaviors.

For our chosen forcing, $F=8$, the steady state solution, $x_m=F$ for
$m=1,\cdots,M$, in (\ref{eq:lorenz}) is linearly unstable. This
instability is associated with unstable dispersive waves characterized by
westward (i.e., in the direction of decreasing $m$) phase velocities and
eastward group velocities. Lorenz and Emanuel
(1998) demonstrated by numerical experiments for $F=8$ and $M=40$ that the
$x$ field is dominated by a wave number 8 structure, and that the system is
chaotic; it has 13 positive Lyapunov exponents, and its Lyapunov dimension
(Kaplan and Yorke 1979) is 27.1. It can be expected that, due to the
eastward group velocities, growing uncertainties in the knowledge of the
model state propagate eastward. A similar process can be observed in
operational numerical weather forecasts, where dispersive short
(longitudinal wave number 6-9) Rossby waves, generated by baroclinic
instabilities, play a key role in the eastward propagation of uncertainties
(e.g., Persson 2000; Szunyogh et al. 2002; and Zimin et al. 2003).

We carried out experiments with three different size systems ($M=i\times 40,$
$i=1,2,3$) and found that increasing the number of variables did not change the
wavelength, i.e. the $x$ fields were dominated by  wavenumber $i\times 8$
structures.

\subsection{Rms analysis error}
The 40-variable version of the L96 model was also used by Whitaker and Hamill (2002)
to validate their ensemble square root filter (EnSRF) approach. In designing our
OSSE's we follow their approach of first generating the `true state', $x^t_m(t)$,
$m=1,\cdots,M$, by a long (40,000 time-step) model integration; then first creating
`observations' of all model variables at each time step by adding uncorrelated
normally distributed random noise with unit variance to the `true state' (i.e.,
$\mathbf{R}_{m}=\mathbf{I}$). (The rms random observational noise variance of 1.00
is to be compared with the value 3.61 of the time mean rms deviation of solutions,
$x_m(t)$, of (\ref{eq:lorenz}) from their mean.) We found that our results were the
same for Gaussian noise and for truncated Gaussian noise (we truncated at three
standard deviations). The effect of reduced observational networks is studied by
removing observations one by one, starting from the full network, at randomly
selected locations. The reduced observational networks are fixed for all
experiments. That is, the difference between a network with $O$ observations and
another with $O+1$ observations is that there is a fixed location at which only the
latter takes observations.

The observations are assimilated at each time step, and
the accuracy of the analysis is measured by the time mean of the rms error,
\begin{equation}
E=\Big({\frac{1}{M}\sum_{m=1}^{M}(\bar{x}^a_m-x^t_m})^2\Big)^{1/2}.
\end{equation}

\subsection{Reference data assimilation schemes}
In order to the assess the skill of our data assimilation scheme in shadowing
the true state, we considered three alternative schemes for comparison.

\subsubsection{Full Kalman filter}
For the sake of comparison with our local ensemble Kalman filter results, we first
establish a standard that can be regarded as the best achievable ensemble Kalman
filter result that could be obtained given that computer resources placed no
constraint on computations of the analysis. (In contrast with operational weather
prediction, for our simple $M$-variable Lorenz model, this is indeed the case.) For
this purpose, we considered the state
$\mathbf{x}(t)=\big(x_1(t),x_2(t),\cdots,x_{M}(t)\big)$ on the entire domain rather
than on a local patch. Then several Kalman filter runs were carried out with
different numbers of ensemble members. In these integrations, full ($k'$) rank
estimates of the covariance matrices were considered and the ensemble perturbations
were updated using (\ref{eq:potter4}), (\ref{eq:Y3}), and (\ref{eq:delta}) of
Appendix~B. (see Section~\ref{sec:sol_of}).

We found that stable cycling of the full ensemble Kalman filter requires increasing
variance inflation  when the number of observations is reduced, even if several
hundred ensemble members are used (e.g., the assimilation of 21 observations
required 2\% variance inflation). This suggests that  variance inflation is needed,
not to compensate for sampling errors, but to correct for variance lost to nonlinear
effects.

It can be seen that by increasing the number of ensemble members the time mean of
$E$ converges to 0.20 regardless of $M$ (figure~\ref{fig:3}). The only difference
between the different size systems (characterized by different values of $M$) is
that more ensemble members are required to reach the minimum value for the larger
systems. We refer to 0.2 as the \emph{``optimal'' error}, and we regard it as a
comparison standard for our local Kalman filter method. (However, we note that it is
not truly optimal since Kalman filters are rigorously optimal only for linear
dynamics.)

\subsubsection{Conventional method}
We designed another comparison scheme that we call the \emph{conventional method},
to obtain an estimate of the analysis error that can be expected from a procedure
analogous to a 3D-Var scheme adapted to the L96 model. In this scheme, only the best
estimate of the true state is sought (not an ensemble of analyses) using a constant
estimate of the background error covariance matrix that does not change with time or
position. This background error covariance matrix was determined by an iterative
process. In the first step, the background error covariance matrices from the full
Kalman filter were averaged over all locations and time steps to obtain a first
estimate. Then, a time series of the true background error vector
$\mathbf{b}_m=\mathbf{x}_m^t-\bar{\mathbf{x}}_m^b$ was generated and used to obtain
an estimate of the background error covariance matrix for the next iteration step.
This step was repeated until the estimated background error covariance matrix
converged, where the convergence was measured by the Frobenius matrix norm. We found
that this procedure was always convergent when all variables were observed. The
estimate obtained this way is not necessarily optimal in the sense of providing the
smallest possible analysis error of any constant background error matrix, but it
 has the desirable feature that the background error statistics are
correctly estimated by the analysis scheme. This is a big advantage compared to the
operational schemes, for which the estimate of the background error covariance
matrix has to be computed by rather ad hoc techniques, since the true state, and
therefore the true background error statistics, are not known. Thus, it might be
assumed that our ``conventional method" provides an estimate of the analysis error
that is of good accuracy as compared to analogous operational schemes.

For reduced observational networks ($O<M$), the background error covariance matrix
was determined by starting the iteration from the background error covariance matrix
for $O+1$. It was found that, when more than a few observations (more than 6 for
$M=40$) were removed, our iterative determination of background error covariance
matrices started to diverge after an initial phase of convergence. This probably
occurs because the background error becomes inhomogeneous, due to the inhomogeneous
observing network, and the average background error underestimates the error at the
locations where the background error is larger than average. This leads to a further
increase of the background error at some locations, resulting in an overall
underestimation of the background error. This highlights an important limitation of
the schemes based on a static estimate of the background error covariance matrix:
The data assimilation scheme must overestimate the average background error in order
to prevent the large local background errors from further growth. Keeping this in
mind, we chose that member of our iteration scheme that provided the smallest
analysis error.

\subsubsection{Direct insertion}
We now give a third standard designed to decide whether the data assimilation
schemes provide any useful information compared to an inexpensive and simple scheme,
not requiring matrix operations. This scheme, called \emph{direct insertion},
updates the state estimate by replacing the background with the observations, where
observations are available, and leaving the background unchanged, where there are no
observations.

\subsection{Implementation of the Local ensemble Kalman filter}
We now describe the implementation of our method on the L96 model. From
(\ref{eq:analysis2}) we know that for our OSSE's ($\mathbf{R}_m$ is the $O\times O$
unit matrix), the analysis error covariance matrix is $\hat{\mathbf{P}}_m^a=
\Big[\big(\hat{\mathbf{P}}_m^b\big)^{-1}+\mathbf{Q}_m^T\mathbf{H}^T\mathbf{H}\mathbf{Q}_m
\big]^{-1}$, where
$\mathbf{H}^T\mathbf{H}=\mbox{diag}[\sigma_1,\sigma_2,\ldots,\sigma_m,\ldots,\sigma_M]$
and $\sigma_m$=1 if there is an observation at grid point $m$ and is zero otherwise.
(Here the local regions are labelled by a single subscript $m$ (rather than $mn$ as
used in Section~2.4) corresponding to the one dimensional spatial variable $m$ in
(\ref{eq:lorenz}).) [When observations are at all $M$ grid points
$\mathbf{H}^T\mathbf{H}=\mathbf{I}$, and since $\mathbf{Q}_m^T \mathbf{Q}_m$
commutes with $\hat{\mathbf{P}}^b_m$, we have that $\hat{\mathbf{P}}^b_m$ and
$\hat{\mathbf{P}}^a_m$ commute.  Thus, when every point is observed,
(\ref{eq:Solution1}), (\ref{eq:Solution2}), and (\ref{eq:Solution3}) are identical,
and solution 1, 2, and 3 are the same.] We implement this solution using
(\ref{eq:potter4}), (\ref{eq:Y3}) and (\ref{eq:delta}) of Appendix~B.

In our experiments, the local analysis covariance matrix is computed by
(\ref{eq:an2alt}) and the local analysis is obtained by (\ref{eq:analysis}). The
analysis ensemble is updated by (\ref{eq:potter4}), (\ref{eq:Y3}) and
(\ref{eq:delta}) of Appendix~B, and the variance of the background ensemble is
increased by a factor of $1+\varepsilon$ in each step using the enhanced variance
inflation algorithm (see Appendix~C for detail). The final analysis at each point
$m$ is computed by averaging $(2l'+1)=5$ local analyses (see figure~\ref{fig:2}). We
found, by numerical experimentation, that achieving the same accuracy requires fewer
ensemble members using averaging instead of simply inserting the center point.
Choosing $l=6$, $k=k'$, and $\varepsilon=0.12$, values that for $M=40$ gave the
lowest mean error (0.2), we found that the mean error does not change with
increasing $M$.

Figure~\ref{fig:4} shows, that when the ensemble has at least eight members, the
analysis error settles at the level (0.2) of the "optimal" scheme, independently of
the number of variables. This is roughly consistent with the supposition of an
effective correlation length for the dynamics that is less than $M$. Thus our method
appears to be effective on large systems of this type. Moreover, the
(non-parallelized) analysis computational time scales linearly with the number of
local regions (i.e., with $M$). This favorable scaling is to be expected, since the
analysis computation size in each local region is independent of $M$.

We note, that the aforementioned scaling property of the local Kalman filter is in
contrast to the behavior of the full Kalman filter, which requires many more
members, and also an increasing number of members for an increasing number of
variables, to achieve the "optimal" precision. This demonstrates the potential
superiority of the local Kalman filter in terms of computational efficiency when
applied to large systems. On the other hand, it also means that, since the minimum
error was independent of $M$, it suffices to use the smallest, 40-variable, system
for further experimentation.

\subsection{Comparison of the data assimilation schemes}
The four data assimilation schemes (local ensemble Kalman Filter, full Kalman
filter, conventional method, and direct insertion) were compared for different
numbers of observations (figure~\ref{fig:5}). The two Kalman filter schemes give
almost identical error results, although the full Kalman filter has a very small
advantage. The two Kalman filter schemes and the conventional data assimilation
scheme are always more accurate than direct insertion, indicating that they are
always able to retrieve nontrivial, useful information about the true state. The two
Kalman filter schemes, in addition, have a growing advantage over the conventional
scheme as the number of observations is decreased. This shows that, as the
observational network and the background error become more inhomogeneous, the
adaptive nature of the background error covariance matrix in the Kalman filters
leads to a growing advantage over the static scheme.

The above numerical experimentation results provide a guide for making good
parameter choices in the case of the L96 model.  In future applications to actual
weather models, choices for the analysis parameters might similarly be determined by
experimentation, but it would also be useful to obtain some guides for initial
guesses of good parameter choices.

\subsection{Sensitivity to the free parameters}
The free parameters of our scheme are the dimensionality of the local regions (which
is $2l+1$), the rank of the covariance matrices ($k$), and the coefficient
($\varepsilon$) in the enhanced variance inflation algorithm. These parameters have
been fixed so far. In what follows, the sensitivity of the data assimilation scheme
to the tunable free parameters is investigated by numerical experiments ($k'$ is
held fixed at $k'=9$). In these experiments, our `true state' and observations are
generated in the same way as in Whitaker and Hamill (2002) ($O=M$). Also, we use the
same ensemble size as Whitaker and Hamill ($k'+1=10$). Hence our analysis error
results and theirs can be directly compared.

In the first experiment the variance inflation coefficient is constant,
$\varepsilon=0.012$, while the dimension of the local vectors ($2l+1$) and the rank
($k$) of the background covariance matrix are varied. The results are shown in
Table~1. The scheme seems to be stable and accurate for a wide range of parameters.
The optimal size local region consists of $2l+1=11,13$ grid points, at which  rank
$k=5,6,7,8,9$ estimates of the background covariance matrix provide similarly
accurate analyses. Moreover, rank 3 and 4 estimates lead to surprisingly accurate
analyses for the smaller size ($2l+1=5,7,9$) local regions. This indicates that the
background uncertainty in a local region at a given time ($\hat{\mathbf{P}}_m^b$)
can be well approximated in a low ($k$) dimensional linear space. Our premise, that
the dimension of this space can be significantly lower than the number of ensemble
members ($k'+1$) needed to evolve the uncertainty, proved to be correct for the L96
model. (We note that the  local dimensionality $k$ is also much smaller than the
"global" Lyapunov-dimension, 27.1, of the system). On the practical side, this
result suggests that, at least for the L96 model, the efficiency of the analysis
scheme can be significantly improved by using ranks that are smaller than the
dimension of the local vectors and the number of ensemble members. We note that our
best results are at least as good as the best results published in Whitaker and
Hamill (2002) and attain the optimal value (0.20) from section~5.3.

In the second experiment, the dimension of the local regions is constant
($2l+1=13$), while the rank and the variance inflation coefficient are
varying. The results are shown in Table~2. While lower rank estimates of
the background error covariance matrix require somewhat stronger variance
inflation, the results are not sensitive to the choice of $\varepsilon$
once it is larger than a critical value. (By critical value we mean the
smallest $\varepsilon$ that provides the optimal error).

The second experiment was then repeated by using the  regular variance
inflation of Anderson and Anderson (1999) and Whitaker and Hamill (2002). In the regular
variance inflation, all background ensemble perturbations are multiplied
by $r=1+\delta$, where $\delta$ is small, $1\gg \delta > 0$. This inflation
strategy increases the total variance in the background ensemble by a
factor of $(1+\Delta)=1+\delta^2+2\delta \approx 1+ 2\delta$.
It can be seen from Table~3 that, except for $k=4$, the critical value of
$\varepsilon$ is less than half of the critical value of $\Delta$. The main
difference between the two inflation schemes is that the enhanced scheme
inflates the dominant eigendirections of the background covariance matrix
less aggressively, and the least dominant eigendirections more
aggressively. The numerical results suggest that this feature of the
scheme is beneficial, indicating that the ensemble-based estimate of the
background error is more reliable in the more unstable directions than in
the other directions. This is also well illustrated by the quantitative
results shown in Figure~\ref{fig:6}. To explain this figure, we define
the true
background error, $\mathbf{b}_m=\mathbf{x}_m^t -\bar{\mathbf{x}}_m^b$ by
the difference between the truth, $\mathbf{x}^t_m$ and the background
mean, $\bar{\mathbf{x}}_m^b$. We also define
$b_m^{(j)}=\mathbf{b}^T_m \mathbf{u}_m^{(j)}$, the
component of $\mathbf{b}_m$ along the semi-axis of the probability
ellipsoid, corresponding to the $j$th largest
eigenvalue, $\lambda_m^{(j)}$ of $\mathbf{P}_m^b$, where $j=1,2,\cdots,k$.
(The case $k=2$ is illustrated in Figure~\ref{fig:7}.) For an ensemble
that correctly estimates the uncertainty in each basis direction,
the time means of
\begin{equation}
d_m^{(j)}=\big({b_m^{(j)}}^2/\lambda_m^{(j)}\big)^{1/2},\qquad
j=1,2,\cdots,k,
\end{equation}
should be close to one. When, for a given $j$, the ratio $d_m^{(j)}$ is smaller than
one, the ensemble tends to overestimate the distance between the truth and the
background in the $\mathbf{u}_m^{(j)}$ direction. When $d_m^{(j)}$ is larger than
one, the ensemble underestimates this distance. Figure~\ref{fig:6} shows that with
the enhanced variance inflation the behavior of the ensemble is much better than
with the regular variance inflation.  This is especially true for the less dominant
eigendirections, for which the ensemble with regular variance inflation
significantly (by about a factor of 6) underestimates the distance between the truth
and the mean background. We found that $\|\mathbf{b}_m\|^2-\sum_{j=1}^9
{b_m^{(j)}}^2$, the true background variance unexplained by the directions,
$\mathbf{u}_m^{(j)}$, $j=1,2,\cdots,9$; is about 3\% of the true total variance
($\|\mathbf{b}_m\|^2$) for all four cases shown in Figure~\ref{fig:6}. Thus the
results indicate that the superior performance of the enhanced variance inflation is
due to the better distribution of the variance between the resolved directions. We
note that this advantage of the enhanced variance inflation could not be exploited
if the analysis was not done in $\mathbb{S}_{mn}$ introduced in Section~2.  Whether
the advantage found for our enhanced variance inflation scheme carries over from the
L96 model to a more realistic situation remains to be determined.

An interesting feature is the anomalously large error value of 0.29 at
$\Delta=0.036$, $k=8$ in Table~3. An inspection of the data revealed that the higher
time average is associated with a sudden and short-lived high amplitude spike in the
rms analysis error. A further analysis of the problem revealed that spikes occur
very rarely and they usually have small amplitude (smaller than 1). On rare
occasions, however, the spikes can have large amplitude (sometimes larger than 5),
and they can last a few thousand time steps. This phenomenon is illustrated by
Figure~\ref{fig:8}, in which the first large spike occurs after more than 162,000
time steps (equivalent to about 104 years, assuming that one time step is equivalent
to 6 hours) and lasts about 12,000 time steps (12 years in real time), and a second
large spike develops 230,000 time steps (146 years in real time) later, which lasts
for 3000 steps (2 years). The severity of this problem was studied by carrying out
several long term integrations with different combinations of the tunable
parameters. An interesting feature is that the spikes do not destroy the overall
stability of the cycle; the large errors always disappear after a finite time and
the mean error is smaller than 0.3. (For the case shown in Figure~\ref{fig:8} the
time mean error is 0.23). Spikes occur regardless of the size of the local region,
and the type of the variance inflation scheme. They become less frequent, however,
as the rank and the variance inflation are increased. In particular, no spikes were
observed for $\varepsilon \ge 0.022$. This suggests that the easiest way to prevent
the occurrence of spikes is to choose a large enough variance inflation coefficient.

All results shown so far were obtained using (\ref{eq:delta}) to generate the
analysis ensemble, $\mathbf{X}_{m}^a$. This scheme results in analysis perturbations
of the form $\delta\mathbf{x}_m^{a(i)}= \delta\mathbf{x}_m^{a(i)
(\parallel)}+\delta\mathbf{x}_m^{a(i)(\perp)}$ as required by
(\ref{eq:eq4_1})-(\ref{eq:perpcomp}). In order to test the importance of including
the small $\delta \mathbf{x}_m^{a(i)(\perp)}= \mathbf{x}_m^{b(i)(\perp)}$ component,
the first experiment was repeated by using solution~1 [(\ref{eq:Solution1})] for
$\mathbf{Z}_{m}$ and $\delta \mathbf{x}_m^{a(i)(\perp)}=\mathbf{0}$ instead of
(\ref{eq:perpcomp}). (Using solution~1 and (\ref{eq:perpcomp}) would give the same
result as (\ref{eq:delta}) for our choice of $\mathbf{R}_{mn}=\mathbf{I}$.) This
modified scheme, restricting the analysis perturbations to the $k$ dimensional space
$\mathbb{S}_m$, is clearly inferior (compare Tables~2~and~4 and Tables~3~and~5).
More precisely, the constrained scheme provides stable analysis cycles only if both
$k$ and $\varepsilon$ are relatively large. This is not unexpected, since setting
the component $\delta \mathbf{x}_{m}^{a(i)(\perp)}$ to zero artificially reduces the
total variance, $\| \delta \mathbf{x}_m^{a(i)} \|^2$. Increasing $k$ decreases the
reduction in the total variance, while increasing $\varepsilon$ compensates for an
increasing part of the lost variance. Also, the constrained scheme is more stable
when the enhanced variance inflation is used, indicating that correcting the
distribution of the variance is not less important than increasing the total
variance.

\subsection{Discussion of the significance of the numerical experiments}
Finally, we reemphasize that the significance of the results of all our numerical
experiments on the toy model (\ref{eq:lorenz}) is limited. Many important factors of
real weather forecasting are not represented (e.g., model error), and very idealized
conditions are assumed (e.g., known, normal, uncorrelated, unbiased, observation
errors, and no ``subgrid scale" stochastic-like input to the evolution of the
``truth" state). On the other hand, it is also probably reasonable to assume that,
if our assimilation procedure gave unfavorable results for our idealized toy model
situation, then the scheme would also be unlikely to be effective in the real case.
Thus, one can view the good results obtained with our assimilation scheme in these
numerical experiments as necessary, but certainly not sufficient, for future
successful performance in a real situation.

\section{Summary and conclusions}

In this paper, we have introduced a local method for assimilating atmospheric data
to determine best-guess current atmospheric states. The main steps in our method are
the following.
\begin{itemize}
\item The global analysis ensemble members are advanced by the atmospheric model to
obtain the global background ensemble at the next analysis time. \item In each local
region, each background ensemble member's perturbation from the ensemble mean is
used to construct a `local vector'. \item Each of the local vectors in the ensemble
is projected onto the local low dimensional subspace. \item The observations are
assimilated in each local region. \item The local analyses are used to determine the
global analysis and an ensemble of  global analysis states. \item The cycle is then
repeated.
\end{itemize}

Numerical tests of the method using the Lorenz model,
(\ref{eq:lorenz}), have been performed. These tests indicate that the method
is potentially very effective in assimilating data.
Other potential favorable features of our method are that only low
dimensional matrix operations are required, and that the analyses
in each of the local regions are independent, suggesting the use of
efficient parallel computation. These features should make possible
fast data assimilation in operational settings. This is supported by
preliminary work (not reported in this paper) in which we have implemented
 our method on the T62, 28-level version of the National Centers for
Environmental Prediction Medium Range Forecasting Model (NCEP MRF). The
assimilation of a total number of $1.5 \times 10^6$ observations (including wind,
temperature, and surface pressure observations) at $k'=k=39$ and $2l+1=9$ takes
about 6 minutes CPU time on 40 2.8 GHz Xeon processors.

\section {\Large Acknowledgments} \noindent This work was supported by the W.~M.~Keck
Foundation, a James~S. McDonnell 21st Century Research Award, by the Office of Naval
Research (Physics), by the Army Research Office, and by the National Science
Foundation (Grants \#0104087 and PHYS 0098632).

\section*{Appendix A: Global continuity of matrix square roots}
\appendix

Not all matrix square root definitions yield global continuity. One particular
important mechanism for non-global-continuity of matrix square roots is that the
eigenvectors of a globally continuous, symmetric, non-negative matrix,
$\mathbf{M}(\mathbf{r})$, may not be definable in a globally continuous manner. In
particular, for smooth variation of $\mathbf{M}(\mathbf{r})$ in two dimensions, it
can be shown that there will generically be isolated points in space where two of
the eigenvalues of $\mathbf{M}(\mathbf{r})$ are equal. Following previous
terminology in the field of quantum chaos (e.g., Ott 2002), we call such points
``diabolical points'' (e.g., Berry 1983). Assume that two eigenvalues of
$\mathbf{M}(\mathbf{r})$ denoted $\xi_1(\mathbf{r})$ and $\xi_2(\mathbf{r})$, are
equal at the diabolical point $\mathbf{r}=\mathbf{r}_d$, and denote their associated
orthonormal eigenvectors by $\mathbf{v}_1(\mathbf{r})$ and
$\mathbf{v}_2(\mathbf{r})$. Now consider starting at a point $\mathbf{r}_o \neq
\mathbf{r}_d$ and following a continuous path $C$ that encircles $\mathbf{r}_d$ and
returns to $\mathbf{r}_o$. Then it is shown in the paragraph below that, with
continuous variation of $\mathbf{v}_1(\mathbf{r})$ and $\mathbf{v}_2(\mathbf{r})$
along the path, their directions are flipped by $180^\circ$ upon return to
$\mathbf{r}_o$. This presents no contradiction, since orthonormal eigenvectors are
arbitrary up to a change of sign, but it shows that  a specific choice of
$\mathbf{v}_1(\mathbf{r})$ and $\mathbf{v}_2(\mathbf{r})$ cannot be defined in a
globally continuous manner. The positive symmetric square root
$\big(\mathbf{M}(\mathbf{r})\big)^{1/2}$,
\begin{displaymath}
\big(\mathbf{M}(\mathbf{r})\big)^{1/2}=\sum_j \xi_j^{1/2}(\mathbf{r})
\mathbf{v}_j(\mathbf{r})\mathbf{v}_j^T(\mathbf{r}),
\end{displaymath}
is globally continuous because
$\mathbf{v}_j(\mathbf{r})\mathbf{v}_j^T(\mathbf{r})$ returns to itself
upon circuit around a diabolical point, even though
$\mathbf{v}_j(\mathbf{r})$ may flip by $180^\circ$. Thus the solutions for
$\mathbf{Z}_{mn}$ given in (\ref{eq:Solution1}), (\ref{eq:Solution2}),
and (\ref{eq:Solution3})  will be
globally continuous, since positive symmetric square roots are
used. The Cholesky square root will also yield global
continuity. On the other hand, as an example of one of the choices that is
unsatisfactory, the matrix square root choice,
\begin{displaymath}
\sqrt{\mathbf{M}(\mathbf{r})} = \big(\mathbf{M}(\mathbf{r})\big)^{1/2}
\big[\mathbf{v}_1(\mathbf{r})
\mid \mathbf{v}_2(\mathbf{r}) \mid \cdots \big]^T
\end{displaymath}
is clearly not globally continuous if diabolical points are present.

In order to see how the above discussed property of diabolical points
arises, consider the case of a two dimensional matrix,
\begin{equation}
\mathbf{A}= \left( \begin{array}{cc}
\alpha(\mathbf{r}) & \gamma(\mathbf{r}) \\
\gamma(\mathbf{r}) & \beta(\mathbf{r}) \end{array} \right).
\end{equation}
The two eigenvalues of this matrix are
\begin{equation}
\xi_{1,2}(\mathbf{r}) =
\frac{\alpha(\mathbf{r}) + \beta(\mathbf{r})}{2} \pm \bigg[
\Big( \frac{\alpha(\mathbf{r}) - \beta(\mathbf{r})}{2} \Big) +
\gamma^2(\mathbf{r}) \bigg]^{1/2}.
\end{equation}
The eigenvalues are equal when the square root is zero, i.e., when
$\alpha(\mathbf{r})=\beta(\mathbf{r})$ and $\gamma(\mathbf{r})=0$. These equations
represent curves in the two-dimensional $\mathbf{r}$-space, and then, as illustrated
in Figure \ref{fig:10}a, equal eigenvalues $(\xi_1=\xi_2)$ generically occur at
points of intersection of these curves (e.g. the point $\mathbf{r}_d$ shown in the
figure). It suffices to consider the neighborhood of such a diabolical point, and to
use deformed coordinates in which $\gamma$ and $(\alpha - \beta)/2$ serve as axes
(Figure \ref{fig:10}b). Assume that we circuit around the origin of this system on
the circular path $C$ of radius $\rho$, $1 \to 2 \to 3 \to 4 \to 5$ shown in Figure
\ref{fig:10}b. Letting $(\alpha-\beta)/2=\rho \cos \vartheta$, $\gamma=\rho \sin
\vartheta$, this path corresponds to continuous increase of $\vartheta$ from $0$ to
$2 \pi$. On this path one can show that
\begin{equation}
\mathbf{A} - \xi_1 \mathbf{I} = 2 \rho \left( \begin{array}{cc}
\sin \vartheta/2 & 0 \\ 0 & \cos \vartheta/2 \end{array} \right)
\left( \begin{array}{cc} -\sin \vartheta/2 & \cos \vartheta/2 \\
\sin \vartheta/2 & - \cos \vartheta/2 \end{array} \right).
\end{equation}
Thus the normalized eigenvector corresponding to $\xi_1$ is $\mathbf{v}_1= (\cos
\vartheta/2, \sin \vartheta/2)$. As shown in Figure \ref{fig:10}b, continuous
increase of $\vartheta$ from point $1$ to point $5$ results in a $180^0$ flip in the
orientation of $\mathbf{v}_1$. Thus $\mathbf{v}_1$ (and similarly $\mathbf{v}_2$)
cannot be defined in a single-valued globally continuous manner.

\section*{Appendix B: $\mathbf{X}_{mn}^a$ obtained directly from
$\mathbf{Y}_{mn}$} In this Appendix we show that $\mathbf{Y}_{mn}$ as given by
(\ref{eq:e'}) can be used to directly obtain the analysis $\mathbf{X}_{mn}^a
=(k')^{-1/2}\big\{\delta \mathbf{x}_{mn}^{a(1)} \mid \delta \mathbf{x}_{mn}^{a(2)}
\mid \cdots \mid \delta \mathbf{x}_{mn}^{a(k'+1)} \big\}$. In section~4.4, we
discussed a variety of ways to compute a matrix $\mathbf{Y}_{mn}$ to use in
(\ref{eq:rightmult}) to obtain, via (\ref{eq:aeq4_1}), the analysis components,
$\delta \mathbf{x}_{mn}^{a(i)(\parallel)}= \mathbf{Q}_{mn} \delta
\hat{\mathbf{x}}_{mn}^{a(i)}$, in the low dimensional subspace $\mathbb{S}_{mn}$. We
now claim that (\ref{eq:aeq4_1}), (\ref{eq:perpcomp}), (\ref{eq:rightmult}), and
(\ref{eq:e'}) imply that
\begin{equation} \label{eq:delta}
\mathbf{X}_{mn}^a=\mathbf{X}_{mn}^b \mathbf{Y}_{mn}.
\end{equation}
(The crucial difference between (\ref{eq:delta}) and (\ref{eq:rightmult}) is the
absence of the superscribed circumflexes in (\ref{eq:delta})). Then in practice the
columns of (\ref{eq:delta}) can be used directly in (\ref{eq:eq4_1}). First we note
that premultiplication of (\ref{eq:delta}) by $\mathbf{Q}_{mn}^T$ returns
(\ref{eq:rightmult}). Then further premultiplication by $\mathbf{Q}_{mn}$, together
with (\ref{eq:eq21}), yields $\mathbf{\Lambda}_{mn}^{(\parallel)}
\mathbf{x}_{mn}^{(a)} =\mathbf{Q}_{mn} \hat{\mathbf{x}}_{mn}^{(a)}$. This means that
the projection of (\ref{eq:delta}) onto $\mathbb{S}_{mn}$ agrees with
(\ref{eq:aeq4_1}). We verify (\ref{eq:delta}) by showing that its projection into
the complementary spaces $\mathbb{S}_{mn}$ and $\bar{\mathbb{S}}_{mn}$ agree with
the decomposition (\ref{eq:aeq4_1}). It remains to show that the projection of
(\ref{eq:delta}) onto $\bar{\mathbb{S}}_{mn}$ agrees with (\ref{eq:aeq4_1}).
Operating on both sides of (\ref{eq:delta}) with $\mathbf{\Lambda}_{mn}^{(\perp)}$
and using $\hat{\mathbf{X}}_{mn}^{bT}= \mathbf{X}_{mn}^{bT} \mathbf{Q}_{mn}$ in
(\ref{eq:e'}), we have
\begin{equation} \label{eq:square}
\mathbf{\Lambda}_{mn}^{(\perp)} \mathbf{X}_{mn}^a = \mathbf{\Lambda}_{mn}^{(\perp)}
\mathbf{X}_{mn}^b \mathbf{X}_{mn}^{bT} \mathbf{Q}_{mn}
\big(\hat{\mathbf{P}}^b_{mn}\big)^{-1} (\mathbf{Z}_{mn}-\mathbf{I})
\hat{\mathbf{X}}_{mn}^b + \mathbf{\Lambda}_{mn}^{(\perp)} \mathbf{X}_{mn}^b.
\end{equation}
Now we recall from section~2 that $\mathbb{S}_{mn}$ and $\bar{\mathbb{S}}_{mn}$ are
constructed from spanning vectors that are eigenvectors of $\mathbf{P}_{mn}^{b'}$.
Thus $\mathbb{S}_{mn}$ and $\bar{\mathbb{S}}_{mn}$ are invariant under
$\mathbf{P}_{mn}^{b'}$. Since $\mathbf{P}_{mn}^{b'}=\mathbf{X}_{mn}^b
\mathbf{X}_{mn}^{bT}$ (see equation \ref{eq:bcov2}), we have that $\mathbf{X}_{mn}^b
\mathbf{X}_{mn}^{bT}$ commutes with the projection operators
$\mathbf{\Lambda}_{mn}^{(\parallel)}$ and $\mathbf{\Lambda}_{mn}^{(\perp)}$. Thus
\begin{equation} \label{eq:star}
\mathbf{\Lambda}_{mn}^{(\perp)} \mathbf{X}_{mn}^b \mathbf{X}_{mn}^{bT}
\mathbf{Q}_{mn}=\mathbf{X}_{mn}^b \mathbf{X}_{mn}^{bT}
\mathbf{\Lambda}_{mn}^{(\perp)} \mathbf{Q}_{mn}=\mathbf{0},
\end{equation}
where the second equality follows because $\mathbf{Q}_{mn} \hat{\mathbf{w}}$
is in $\mathbb{S}_{mn}$ for any $k$-dimensional column vector $\mathbf{w}$,
thus yielding $\mathbf{\Lambda}_{mn}^{(\perp)} \mathbf{Q}_{mn}=\mathbf{0}$.
From (\ref{eq:square}) and (\ref{eq:star}) we have
$\mathbf{\Lambda}_{mn}^{(\perp)} \mathbf{X}_{mn}^a=
\mathbf{\Lambda}_{mn}^{(\perp)} \mathbf{X}_{mn}^b$ or $\delta
\mathbf{x}_{mn}^{a(i)(\perp)}=\delta \mathbf{x}_{mn}^{b(i)(\perp)}$, as
required by ($\ref{eq:perpcomp}$). This establishes (\ref{eq:delta}). We
find that use of (\ref{eq:delta}) can be potentially advantageous for
efficient parallel implementation of our method. We plan to further
discuss this in a future publication applying our local ensemble
Kalman filter to the operational global model of the National Centers for
Environmental Prediction.

\section*{Appendix C: Enhanced Variance Inflation}
\appendix

In section~4.6 we mentioned the modification of
$\mathbf{P}^a_{mn}$ or $\mathbf{P}^b_{mn}$ to prevent the occurrence of
small eigenvalues in these matrices. Furthermore, we noted the
possibility of an accompanying modification of the corresponding
ensemble perturbations, so as to preserve the relation,
\begin{equation} \label{eq:a1}
\hat{\mathbf{P}}_{mn}= \frac{1}{k'} \sum_{i=1}^{k'+1}
\delta \hat{\mathbf{x}}_{mn}^{(i)} \big( \delta \hat{\mathbf{x}}_{mn}^{(i)}
\big)^T.
\end{equation}
In the above equation we have suppressed the superscript $a$ or $b$ with the
understanding that $(\ref{eq:a1})$ can apply to either the analysis or
background.

We consider the example where $\hat{\mathbf{P}}_{mn}$ is changed to a new
covariance matrix by addition of a small perturbation in the form,
\begin{equation} \label{eq:a3}
\hat{\mathbf{P}}_{mn}^*=\hat{\mathbf{P}}_{mn}+\frac{\varepsilon\Lambda}{k}
\mathbf{I}_k,\qquad \varepsilon>0,
\end{equation}
where $\mathbf{I}_k$ denotes the $k \times k$ unit matrix, and $\Lambda$ is the
trace of $\hat{\mathbf{P}}_{mn}$; i.e., it is the sum of its eigenvalues, and thus
represents the total variance of the ensemble. (The case $k=2$ is illustrated in
Figure~\ref{fig:9}.) Hence (\ref{eq:a3}) increases the total variance by the factor
$(1+\varepsilon)$, where we regard $\varepsilon$ as small, $1 \gg \varepsilon > 0$.
More importantly, for small $\varepsilon$, the additional variance represented in
(\ref{eq:a3}) results in a relatively small change in the largest eigenvalues of
$\hat{\mathbf{P}}_{mn}$, but prevents any eigenvalue from dropping below
($\varepsilon \Lambda/k$), thus effectively providing a floor on the variance in any
eigendirection. Having modified $\hat{\mathbf{P}}_{mn}$ to $\hat{\mathbf{P}}^*_{mn}$
via (\ref{eq:a3}), we now consider the modification of the ensemble perturbations,
$\delta\hat{\mathbf{x}}^{(i)}_{mn}$, to another set of ensemble perturbations,
$\delta\hat{\mathbf{x}}^{(i)*}_{mn}$, with the perturbed covariance,
\begin{equation} \label{eq:a4}
\hat{\mathbf{P}}_{mn}^*=\frac{1}{k'} \sum_{i=1}^{k'+1}
\delta\hat{\mathbf{x}}^{(i)*}_{mn} \big(\delta\hat{\mathbf{x}}^{(i)*}_{mn}\big)^T.
\end{equation}
We use the result of sections~4.2~and~4.4 to choose the $\delta
\hat{\mathbf{x}}^{(i)*}_{mn}$ to minimize the difference with $\delta
\hat{\mathbf{x}}^{(i)}_{mn}$. This result is the same for all metrics
$\mathbf{D}_{mn}$ that commute with $\hat{\mathbf{P}}_{mn}$ (equivalently
$\hat{\mathbf{P}}^*_{mn}$). (Note that the solutions in (\ref{eq:GS1}) are
all the same if $\mathbf{D}_{mn}$, $\hat{\mathbf{P}}^a_{mn}$ and
$\hat{\mathbf{P}}^b_{mn}$ commute.) Adopting this solution for $\delta
\hat{\mathbf{x}}^{(i)*}_{mn}$, we introduce the orthogonal eigenvectors
of $\hat{\mathbf{P}}_{mn}$, which we denote $\mathbf{w}^{(j)}_{mn}$.
The result for $\delta \hat{\mathbf{x}}^{(i)*}_{mn}$ is then
\begin{equation}
\delta \hat{\mathbf{x}}^*_{mn}=\mathbf{Z}^*_{mn} \delta \hat{\mathbf{x}}_{mn},
\end{equation}
where
\begin{equation} \label{eq:a9}
\mathbf{Z}^*_{mn} = \sum_{j=1}^k \xi_{mn}^{(j)}
\mathbf{w}_{mn}^{(j)} \big(\mathbf{w}_{mn}^{(j)}\big)^T
\end{equation}
with
\begin{equation} \label{eq:a9b}
\xi_{mn}^{(j)} = \sqrt{1+\big(\varepsilon \Lambda/k \eta_{mn}^{(j)}\big)},
\end{equation}
and $\eta_{mn}^{(j)}$ is the eigenvalue of $\hat{\mathbf{P}}_{mn}$
corresponding to $\mathbf{w}_{mn}^{(j)}$; that is,
$\hat{\mathbf{P}}_{mn}^{(j)} \mathbf{w}_{mn}^{(j)} =
\eta_{mn}^{(j)} \mathbf{w}_{mn}^{(j)}$.

Recalling that $\hat{\mathbf{P}}^b_{mn}$
is diagonal (see (\ref{eq:eq19})), we see that in the case
$\hat{\mathbf{P}}_{mn}=\hat{\mathbf{P}}^b_{mn}$ (which is employed in
section~5) the $i$th component of the vector $\mathbf{w}^{j}_{mn}$ is
$\delta_{ij}$. Consequently,
for this case, (\ref{eq:a9}) and (\ref{eq:a9b}) imply that
$\mathbf{Z}^*_{mn}$ is diagonal,
\begin{equation}
\mathbf{Z}^*_{mn}=diag(\xi_1, \xi_2,\cdots,\xi_k).
\end{equation}

In the case $\hat{\mathbf{P}}_{mn}=\hat{\mathbf{P}}_{mn}^a$, one could combine
variance inflation and a procedure for obtaining the analysis ensemble
$\big\{\delta \hat{\mathbf{x}}^{a(i)}_{mn}
\big\}$ (e.g., solutions~1,~2,,~or~3 of section~4.2): First inflate
$\hat{\mathbf{P}}^a_{mn}$,
\begin{displaymath}
\hat{\mathbf{P}}^{a*}_{mn}=\hat{\mathbf{P}}^a_{mn} +
\hat{\mathbf{G}}^a_{mn},
\end{displaymath}
where $\hat{\mathbf{G}}^a_{mn}$ is any chosen inflation; and, second, replace
$\hat{\mathbf{P}}^a_{mn}$ by $\hat{\mathbf{P}}^{a*}_{mn}$ in the chosen algorithm
for determining the analysis ensemble. \vspace{1.0cm}

\noindent {\Large References}

\vspace{0.5cm}

\noindent Anderson, J. L., 2001: An ensemble adjustment filter for data
assimilation. \emph{Mon. Wea. Rev.,} {\bf 129,} 2884-2903.

\vspace{0.3cm}
\noindent Anderson, J. L., and S. L. Anderson, 1999: A Monte Carlo
implementation of the nonlinear filtering problem to produce ensemble
assimilations and forecasts. \emph{Mon. Wea. Rev.,} {\bf 127,}
2741-2758.

\vspace{0.3cm}

\noindent Berry, M. V., 1983: Semiclassical mechanics of regular and irregular
motion. In \emph{Chaotic Behavior of Deterministic Systems,} R.~H.~G. Helleman
and G. Ioos, eds., North-Holland, Amsterdam.

\vspace{0.3cm}

\noindent Bierman, G. J., 1977: \emph{Factorization Methods for
Discrete Sequential Estimation}. Mathematics in Science and
Engineering, Academic Press, New York.

\vspace{0.3cm}

\noindent Bishop, C. H., B. J. Etherton, and S. Majumdar, 2001: Adaptive
sampling with the Ensemble Transform Kalman Filter. Part I: Theoretical
aspects. \emph{Mon. Wea. Rev.,} {\bf 129,} 420-436.

\vspace{0.3cm}

\noindent Daley, R., 1991: \emph{Atmospheric data analysis.} Cambridge
University Press, New York.

\vspace{0.3cm}

\noindent Dee, D., S. Cohn, A. Dalcher, and M. Ghil, 1985: An efficient
algorithm for estimating noise covariances in distributed systems.
\emph{IEEE Trans. Automatic Control,} {\bf 30,} 1057-1065.

\vspace{0.3cm}

\noindent Evensen, G., 1994: Sequential data assimilation with a nonlinear
quasi- \linebreak geostrophic model using Monte Carlo methods to
forecast error statistics.
\emph{J. Geophys. Res.,} {\bf 99}(C5), 10 143-10 162.

\vspace{0.3cm}

\noindent Evensen, G., and P. J. van Leeuwen, 1996: Assimilation of Geosat
altimeter data for the Agulhas current using the ensemble Kalman Filter
with a quasi-geostrophic model. \emph{Mon. Wea. Rev.,} {\bf 124,} 85-96.

\vspace{0.3cm}

\noindent Fisher, M., 1998: Development of a simplified Kalman Filter.
ECMWF Research Department Tech. Memo. 260, 16 pp. [Available from European
Centre for Medium-Range Weather Forecasts, Shinfield Park, Reading,
Berkshire, RG2 9AX, United Kingdom.]

\vspace{0.3cm}

\noindent Ghil, M., S. Cohn, J. Tavantzis, K. Bube, and E. Isaacson, 1981:
Applications of estimation theory to numerical weather prediction. In
\emph{Dynamic meteorology: data assimilation methods,} L. Bengtsson, M.
Ghil, and E. Kallen, eds. Springer-Verlag, New York, 139-224.

\vspace{0.3cm}

\noindent Hamill, T. M., and C. Snyder, 2000: A hybrid ensemble
Kalman Filter-3D variational analysis scheme. \emph{Mon. Wea. Rev.,}
{\bf 128,} 2905-2919.

\vspace{0.3cm}

\noindent Hamill, T. M., J. Whitaker, and C. Snyder, 2001:
Distance-dependent filtering of background error covariance estimates in
an Ensemble Kalman Filter. \emph{Mon. Wea. Rev.,} {\bf 129,} 2776-2790.

\vspace{0.3cm}

\noindent Houtekamer, P. L., and H. L. Mitchell, 2001: A sequential ensemble
Kalman Filter for atmospheric data assimilation. \emph{Mon. Wea. Rev.,}
{\bf 129,} 796-811.

\vspace{0.3cm}

\noindent Houtekamer, P. L., and H. L. Mitchell, 1998: Data assimilation using
an ensemble Kalman Filter technique. \emph{Mon. Wea. Rev.,} {\bf 126,}
796-811.

\vspace{0.3cm}
\noindent Houtekamer, P. L., and H. L. Mitchell, 2001: A sequential
ensemble Kalman Filter for atmospheric data assimilation. \emph{Mon.
Wea. Rev.,} {\bf 129,} 123-137.

\vspace{0.3cm}

\noindent Ide, K., P. Courtier, M. Ghil, and A. C. Lorenc, 1997: Unified
notation for data assimilation: Operational, sequential, and variational.
\emph{J. Meteor. Soc. Japan,} {\bf 75}(1B), 181-189.

\vspace{0.3cm}

\noindent Jones, R., 1965: An experiment in nonlinear prediction.
\emph{J. Appl. Meteor.,} {\bf 4,} 701-705.

\vspace{0.3cm}

\noindent Kalman, R., 1960: A new approach to linear filtering and prediction
problems. \emph{Trans. ASME, Ser. D, J. Basic Eng.,} {\bf 82,} 35-45.

\vspace{0.3cm}

\noindent Kalman, R., R. Bucy, 1961: New results in linear filtering and
prediction theory. \emph{Trans. ASME, Ser. D, J. Basic Eng.,} {\bf 83,}
95-108.

\vspace{0.3cm}

\noindent Kalnay, E., 2002: \emph{Atmospheric modeling, data assimilation, and
predictability.} Cambridge University Press, Cambridge, 341 pp.

\vspace{0.3cm}

\noindent Kalnay, E., and Z. Toth, 1994: Removing growing errors in the analysis.
Pre\-prints, \emph{10th AMS Conference on Numerical Weather Prediction,} Portland,
OR, 212-215.

\vspace{0.3cm}

\noindent Keppenne, C., and H. Rienecker, 2002: Initial testing of a massively
parallel ensemble Kalman Filter with the Poseidon Isopycnal Ocean General
Circulation Model. \emph{Mon. Wea. Rev.,} {\bf 130,} 2951-2965.

\vspace{0.3cm}

\noindent Lorenc, A., 1986: Analysis methods for numerical weather prediction.
\emph{Quart. J. Roy. Meteor. Soc.,} {\bf 112,} 1177-1194.

\vspace{0.3cm}

\noindent Lorenz, E. N., K. A Emanuel, 1998: Optimal sites for supplementary
weather observations: Simulation with a small model. \emph{J. Atmos. Sci.,}
{\bf 55,} 399-414.

\vspace{0.3cm}

\noindent Lorenz, E. N., 1996: Predictability: A problem partly solved.
\emph{Proc. Seminar on Predictability,} Vol. 1, ECMWF, Reading, Berkshire,
UK, 1-18.

\vspace{0.3cm}

\noindent Ott, E., 2002: \emph{Chaos in dynamical systems,} (second edition).
Cambridge University Press, Cambridge, chapter 11.

\vspace{0.3cm}

\noindent Patil, D. J., B. R. Hunt, E. Kalnay, J. A. Yorke, and E. Ott, 2001:
Local low dimensionality of atmospheric dynamics. \emph{Phys. Rev. Lett.,}
{\bf 86,} 5878-5881.

\vspace{0.3cm}

\noindent Persson A., 2000: Synoptic-dynamic diagnosis of medium range
weather forecast systems. \emph{Proceedings of the Seminars on Diagnosis
of models and data assimilation systems.} 6-10 September 1999, ECMWF,
Reading, U.K., 123-137.

\vspace{0.3cm}

\noindent Petersen, D., 1973: A comparison of the performance of quasi-optimal
and conventional objective analysis schemes. \emph{J. Appl. Meteor.,}
{\bf 12,} 1093-1101.

\vspace{0.3cm}

\noindent Szunyogh, I., Z. Toth, A. V. Zimin, S. J. Majumdar, and A.
Persson, 2002: Propagation of the effect of targeted observations: The
2000 Winter Storm Reconnaissance Program. \emph{Mon. Wea. Rev.,}
{\bf 130,} 1144-1165.

\vspace{0.3cm}

\noindent Talagrand, O., 1981: A study of the dynamics of four-dimensional
data assimilation. \emph{Tellus}, {\bf 33,} 43-60.

\vspace{0.3cm}

\noindent Tippett, M. K., J. L. Anderson, C. H. Bishop, T. M. Hamill,
and J. S. Whitaker, 2002: Ensemble square-root filters. \emph{Mon. Wea. Rev.,} {\bf 131,} 1485-1490. 

\vspace{0.3cm}

\noindent Toth, Z., and E. Kalnay, 1993: Ensemble forecasting at NMC: The generation
of perturbations. \emph{Bull. Amer. Meteorol. Soc.,} {\bf 74,} 2317-2330.

\vspace{0.3cm}

 \noindent Toth, Z., and E. Kalnay, 1997: Ensemble forecasting at NCEP
and the breeding method. \emph{Mon. Wea. Rev.,} {\bf 127,} 1374-1377.

\vspace{0.3cm}

\noindent Wang, X., and C. H. Bishop, 2002: A comparison of breeding and
Ensemble Transform Kalman Filter ensemble forecast schemes. Preprints,
\emph{Symposium on Observations, Data Assimilation, and Probabilistic
Prediction}, Orlando, FL., Amer. Meteor. Soc., J28-J31.

\vspace{0.3cm}

\noindent Whitaker, J. S., and T. H. Hamill, 2002: Ensemble Data
Assimilation without perturbed observations. \emph{Mon. Wea. Rev.,}
{\bf 130,} 1913-1924.

\vspace{0.3cm}

\noindent Zimin A. V., I. Szunyogh, D. J. Patil, B. R. Hunt, and E. Ott, 2003:
Extracting envelopes of Rossby wave packets. \emph{Mon. Wea. Rev.,} {\bf 131,}
1011-1017.

\newpage

\begin{center}
{\Large Table Captions}
\end{center}
\vspace{0.5cm}
\noindent {\bf Table~1.}
Dependence of the time mean rms error on the box size ($2l+1$)
and the rank ($k$) of the background covariance matrix. The symbol D
stands for time mean rms errors larger than one, which is the rms mean of
the observational errors. The coefficient of the enhanced variance
inflation is $\epsilon=0.012$.

\vspace{0.3cm}

\noindent {\bf Table~2.}
Dependence of the time mean rms error on the coefficient
($\varepsilon$) of the enhanced variance inflation scheme and the rank
($k$) of the background covariance matrix. The meaning of D is the same
as in Table~1. The window size is 13.

\vspace{0.3cm}

\noindent {\bf Table~3.}
Dependence of the rms analysis error on $\Delta$ in
the regular variance inflation scheme and the rank ($k$) of the
background error covariance matrix. The meaning of D is the
same as in Table~1. The window size is 13.

\vspace{0.3cm}

\noindent {\bf Table~4.}
Same as Table~2 except that Solution~1 and $\delta \mathbf{x}_m^{a(i)(\perp)}
=\mathbf{0}$ is used (instead of \ref{eq:delta}) to obtain the analysis
ensemble.

\vspace{0.3cm}

\noindent {\bf Table~5.}
Same as Table~3 except that Solution~1 and $\delta \mathbf{x}_m^{a(i)(\perp)}=
\mathbf{0}$ is used (instead of \ref{eq:delta}) to obtain the analysis
ensemble.

\newpage

\begin{center}
{\Large Figure Captions}
\end{center}

\vspace{0.5cm}

\noindent {\bf New Figure~1.}

Illustration of the Local Ensemble Kalman Filter scheme as given by the six steps
listed in the introduction. The symbols in the figure are as follows:
\begin{itemize}
\item $\mathbf{x}^{a(i)}(\mathbf{r},t)$ = the analysis ensemble fields as a function
of position $\mathbf{r}$ on the globe at time $t$. \item
$\mathbf{x}^{b(i)}_{mn}(\mathbf{r},t)$ = ensemble of background atmospheric state in
local region $mn$. \item $\mathbb{S}_{mn}$ = the local low dimensional subspace in
region $mn$. \item $\hat{\bar{\mathbf{x}}}^a_{mn}(t)$ = the mean analysis state in
$\mathbb{S}_{mn}$. \item $\hat{\mathbf{P}}^a_{mn}(t)$ = the analysis error
covariance matrix in $\mathbb{S}_{mn}$.
\end{itemize}

\vspace{0.3cm}

\noindent {\bf Figure~1.}
Probability ellipsoid for $\mathbf{x}^b_{mn}$.

\vspace{0.3cm}

\noindent {\bf Figure~2.} Illustration of the local region $(l=5)$ with a central
region $(l'=2)$.

\vspace{0.3cm}

\noindent {\bf Figure~3.} The rms error of the full Kalman filter as function of the
number of ensemble members. Shown are the results for $M=40$ (solid line), $M=80$
(dashed line), and $M=120$ (dotted-dashed line).

\vspace{0.3cm}

\noindent {\bf Figure~4.} The rms error of the local ensemble Kalman filter as
function of the number of ensemble members. Shown are the results for $M=40$ (solid
line), $M=80$ (dashed line), and $M=120$ (dotted-dashed line).

\vspace{0.3cm}

\noindent {\bf Figure~5.} The rms error of the different analysis schemes as
function of the number of observations. Shown are the results for the full Kalman
filter [4\% variance inflation] (dashed line), conventional scheme (dashed-dotted
line), direct insertion (solid line with diamonds), and the local ensemble Kalman
filter [3\% variance inflation] (solid line).

\vspace{0.3cm}

\noindent {\bf Figure~6.}
The ratio $d^{(j)}$ at $m=1$ as function of $j$ for two different
values of $\varepsilon$ and $\Delta$.
\vspace{0.3cm}

\noindent {\bf Figure~7.}
Projection of the true background error, $\mathbf{b}_m$, on the main
axes of the probability ellipsoid.
\vspace{0.3cm}

\noindent {\bf Figure~8.}
The long time evolution of the rms analysis error for a set of parameters
that allow spikes to occur.
\vspace{0.3cm}

\noindent {\bf Figure~9.}
The effect of the enhanced variance inflation (equation \ref{eq:a3})
on the probability ellipsoid. For the case
$\hat{\mathbf{P}}_{mn}=\hat{\mathbf{P}}_{mn}^b$, $\eta_{mn}^{(1)}=
\lambda_{mn}^{(1)}$ and $\eta_{mn}^{(2)}=\lambda_{mn}^{(2)}$.
\vspace{0.3cm}

\noindent {\bf Figure~10.}
(a) The curves $\gamma(\mathbf{r})=0$ and
$\alpha(\mathbf{r})=\beta(\mathbf{r})$ cross at a diabolical point
$\mathbf{r}_d$. (b) The eigenvector $\mathbf{v}_1$ flips by $180^0$ on one
circuit around the diabolical point.
\newpage

\begin{table}
\begin{center}
\begin{tabular}{c c|c c c c c c c}
\hline
\hline
&$k$ &  3   &    4   &    5    &   6   &    7   &    8    &    9   \\
$2l+1$&&&&&&&&\\
\hline
 5 && 0.24 & 0.23   &        &      &     &       &    \\
 7  &&  0.22  &  0.22 &   0.21  &  0.22  &       &       &     \\
 9  &&  0.22  &  0.21 &   0.21  &  0.21  &  0.21  &  0.21 &   \\
11  &&  D  & D  &   0.20  &  0.20  &  0.20  &  0.20 &   0.20 \\
13   && D & D  &   0.20  &  0.20  &  0.20  &  0.20  &  0.20\\
15   && D  & D  &  D  &  0.22  &  0.20  &  0.2  &  0.20 \\
\hline
\hline
\end{tabular}
\caption{Dependence of the time mean rms error on the box size ($2l+1$)
and the rank ($k$) of the background covariance matrix. The symbol D
stands for time mean rms errors larger than one, which is the rms mean of
the observational errors. The coefficient of the enhanced variance
inflation is $\epsilon=0.012$.}
\label{table1}
\end{center}
\end{table}

\clearpage

%%%%%%%%%%%%%%%%%%%%%%%%%%%%%end of first table%%%%%%%%%%%%%%%%%%%%%%%%%%%%%
\begin{table}
\begin{center}
\begin{tabular}{c c|c c c c c c}
\hline
\hline
&$k$ &    4   &    5    &   6   &    7   &    8    &    9    \\
$\epsilon$&&&&&&&\\
\hline
 0.008 && D  &  D  & 0.44 & 0.20 &  0.20 &  0.20\\
 0.010 && D  &  D  & 0.2 & 0.2 &  0.20 &  0.20 \\
 0.012  &&D  &  0.20 & 0.20 & 0.20 &  0.20 &  0.20\\
 0.014 && D  &  0.20 & 0.20 & 0.20 &  0.20 &  0.20  \\
 0.016 && D  &  0.20 & 0.20 & 0.20  & 0.20 &  0.20 \\
 0.018 && D  &  0.20 & 0.20 & 0.20  & 0.20 &  0.20 \\
 0.020 && 0.21 & 0.20 & 0.20 & 0.20 &  0.20 &  0.20\\
\hline
\hline
\end{tabular}
\caption{Dependence of the time mean rms error on the coefficient
($\varepsilon$) of the enhanced variance inflation scheme and the rank
($k$) of the background covariance matrix. The meaning of D is the same
as in Table~1. The window size is 13.}
\label{table2}
\end{center}
\end{table}

\clearpage

\begin{table}
\begin{center}
\begin{tabular}{c c|c c c c c c}
\hline
\hline
&$k$ &    4   &    5    &   6   &    7   &    8    &    9    \\
$\Delta$&&&&&&&\\
\hline
0.020 &&  D  &    D    &  0.50    &  0.30  &   D    &  D  \\
 0.024 &&  D &   0.87 &   0.42  &  0.21 &   0.21  &  0.21 \\
 0.028 &&  0.21 &   0.36 &   0.20  &  0.20 &   0.20  &  0.20\\
 0.032 &&  0.20 &   0.20  &  0.20  &  0.20  &  0.20  &  0.20 \\
 0.036  && 0.20 &   0.20 &   0.20  &  0.20  &  0.29  &  0.20 \\
 0.040  && 0.20 &   0.20 &   0.20   & 0.20  &  0.20  &  0.20 \\
 0.044  && 0.20 &   0.20 &   0.20   & 0.20  &  0.20 &   0.20 \\
 0.048 &&  0.20 &   0.20 &   0.20   & 0.20  &  0.20 &   0.20 \\
\hline
\hline
\end{tabular}
\caption{Dependence of the rms analysis error on $\Delta$ in
the regular variance inflation scheme and the rank ($k$) of the
background error covariance matrix. The meaning of D is the
same as in Table~1. The window size is 13.}
\end{center}
\label{table3}
\end{table}

\clearpage

\begin{table}
\begin{center}
\begin{tabular}{c c|c c c c c c}
\hline
\hline
&$k$ &    4   &    5    &   6   &    7   &    8    &    9    \\
inflation coefficient $\epsilon$&&&&&&&\\
\hline
 0.010 &&  D &  D  & D     &  0.41  &  0.20  &  0.20  \\
 0.012 &&  D  & D  & D     &  0.27  &  0.20  &  0.20 \\
 0.014 &&  D  & D  & D     &  0.21  &  0.20  &  0.20   \\
 0.016 &&  D &  D  & D     &  0.21  &  0.20  &  0.20  \\
 0.018 &&  D &  D  &   0.46  &  0.21  &  0.20  &  0.20  \\
 0.020 &&  D  & D  &   0.28  &  0.21  &  0.20  &  0.20 \\
 0.022  && D &  D  &   0.23  &  0.21  &  0.20  &  0.21 \\
 0.024  && D &  D  &   0.22  &  0.21  &  0.21  &  0.21 \\
\hline
\hline
\end{tabular}
\caption{Same as Table~2 except that Solution~1 and $\delta
\mathbf{x}_m^{a(i)(\perp)}=\mathbf{0}$ is used (instead of \ref{eq:delta}) to
obtain the analysis ensemble.}
\label{table4}
\end{center}
\end{table}

\clearpage

\begin{table}
\begin{center}
\begin{tabular}{c c|c c c c c c}
\hline
\hline
&k &    4   &    5    &   6   &    7   &    8    &    9    \\
$\Delta$&&&&&&&\\
\hline
0.020 &&  D  &  D  &  D  & D  &  D  &  D  \\
 0.024 &&  D &  D  &  D  & D  &  0.75  &  0.21 \\
 0.028 &&  D &  D  &  D  & D  &  0.22  &  0.21\\
 0.032 &&  D &  D  &  D  & D  &  D  &  0.20 \\
 0.036  && D &  D  &  D  & D  &  0.21  &  0.25 \\
 0.040  && D &  D  &  D  & 0.22 &  0.21  &  0.20 \\
 0.044  && D &  D  &  D  & D  &  0.20 &   0.20 \\
 0.048 &&  D &  D  &  D  & 0.25 &  0.20 &   0.20 \\
\hline
\hline
\end{tabular}
\caption{Same as Table~3 except that Solution~1 and $\delta
\mathbf{x}_m^{a(i)(\perp)}=\mathbf{0}$ is used (instead of \ref{eq:delta}) to
obtain the analysis ensemble.}
\label{table5}
\end{center}
\end{table}

\clearpage

\begin{figure}
\begin{center}
\includegraphics[width=1.0\textwidth]{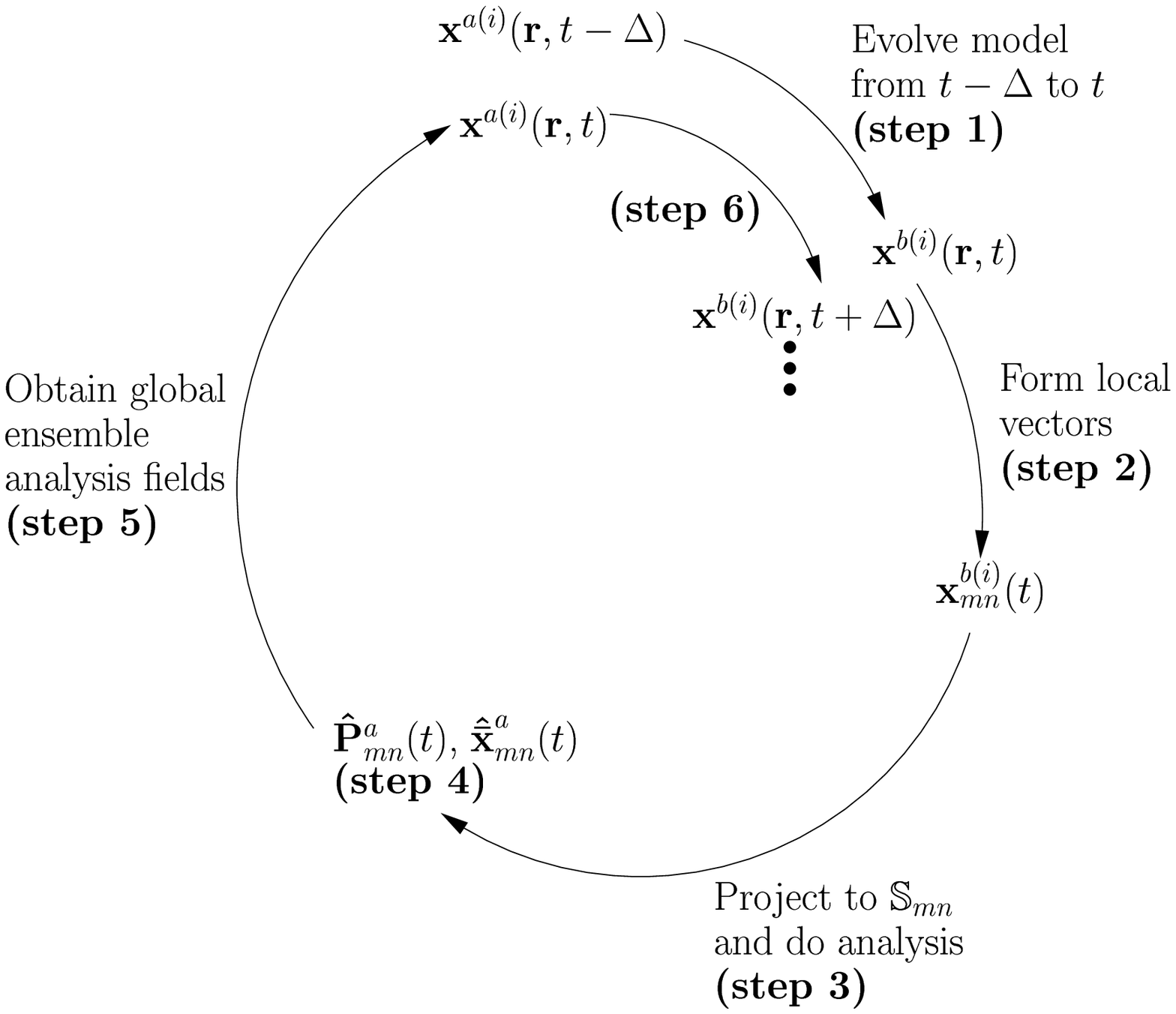}
\parbox{40cm}{The symbols in the figure are as
follows:
\begin{itemize} \item $\mathbf{x}^{a(i)}(\mathbf{r},t)$ = the analysis
ensemble fields as a function of position $\mathbf{r}$ on the globe at time $t$.
\item $\mathbf{x}^{b(i)}_{mn}(\mathbf{r},t)$ = ensemble of background atmospheric
state in local region $mn$. \item $\mathbb{S}_{mn}$ = the local low dimensional
subspace in region $mn$. \item $\hat{\bar{\mathbf{x}}}^a_{mn}(t)$ = the mean
analysis state in $\mathbb{S}_{mn}$. \item $\hat{\mathbf{P}}^a_{mn}(t)$ = the
analysis error covariance matrix in $\mathbb{S}_{mn}$.
\end{itemize}}
\caption[summary]{Illustration of the Local Ensemble Kalman Filter scheme as given
by the six steps listed in the introduction.
 } \label{fig:1_a}
\end{center}
\end{figure}

\begin{figure}
\begin{center}
\includegraphics[width=1.0\textwidth]{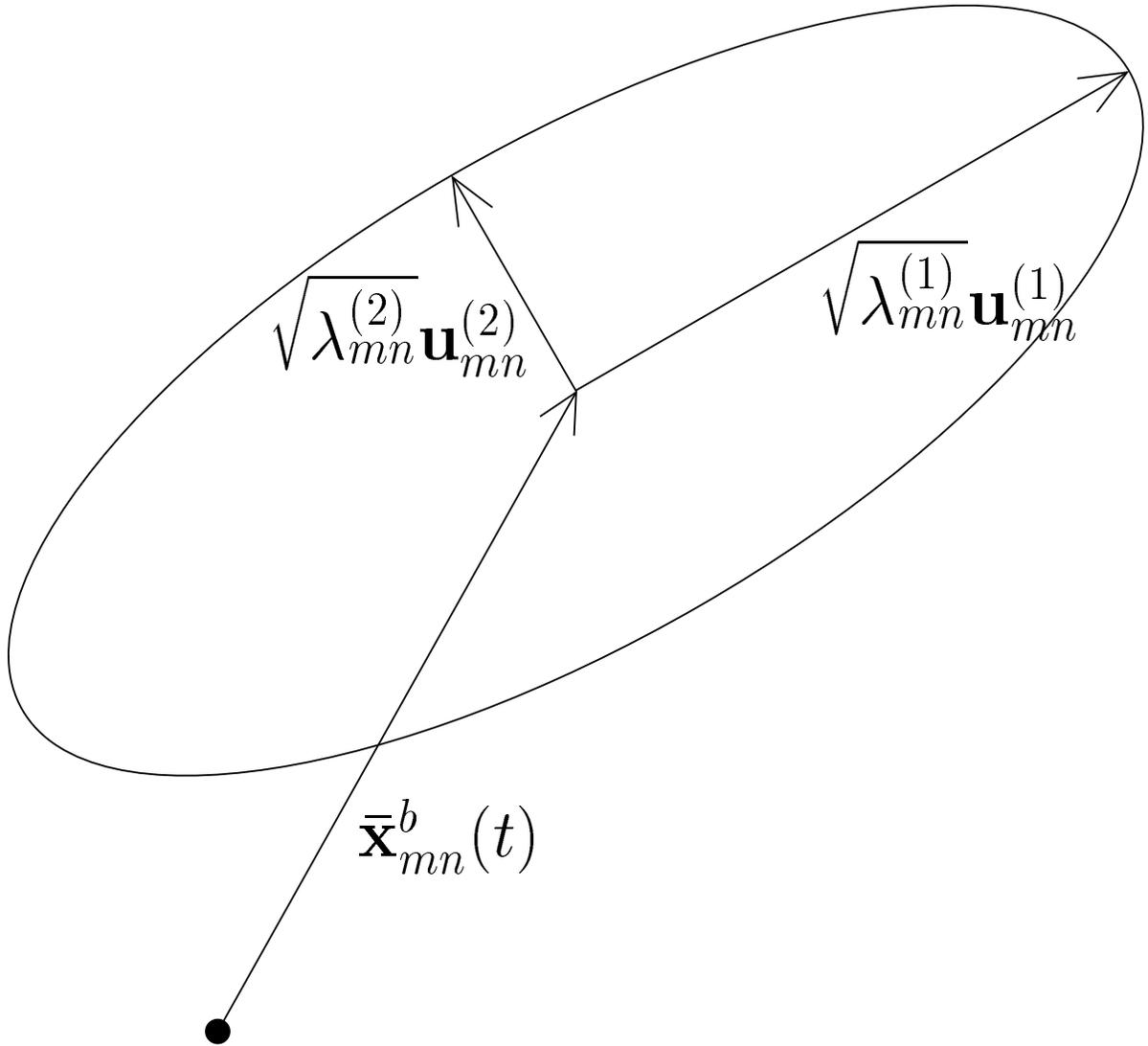}
\caption{Probability ellipsoid for $\mathbf{x}^b_{mn}$.}
\label{fig:1}
\end{center}
\end{figure}

\clearpage

\begin{figure}
\begin{center}
\includegraphics[width=1.0\textwidth]{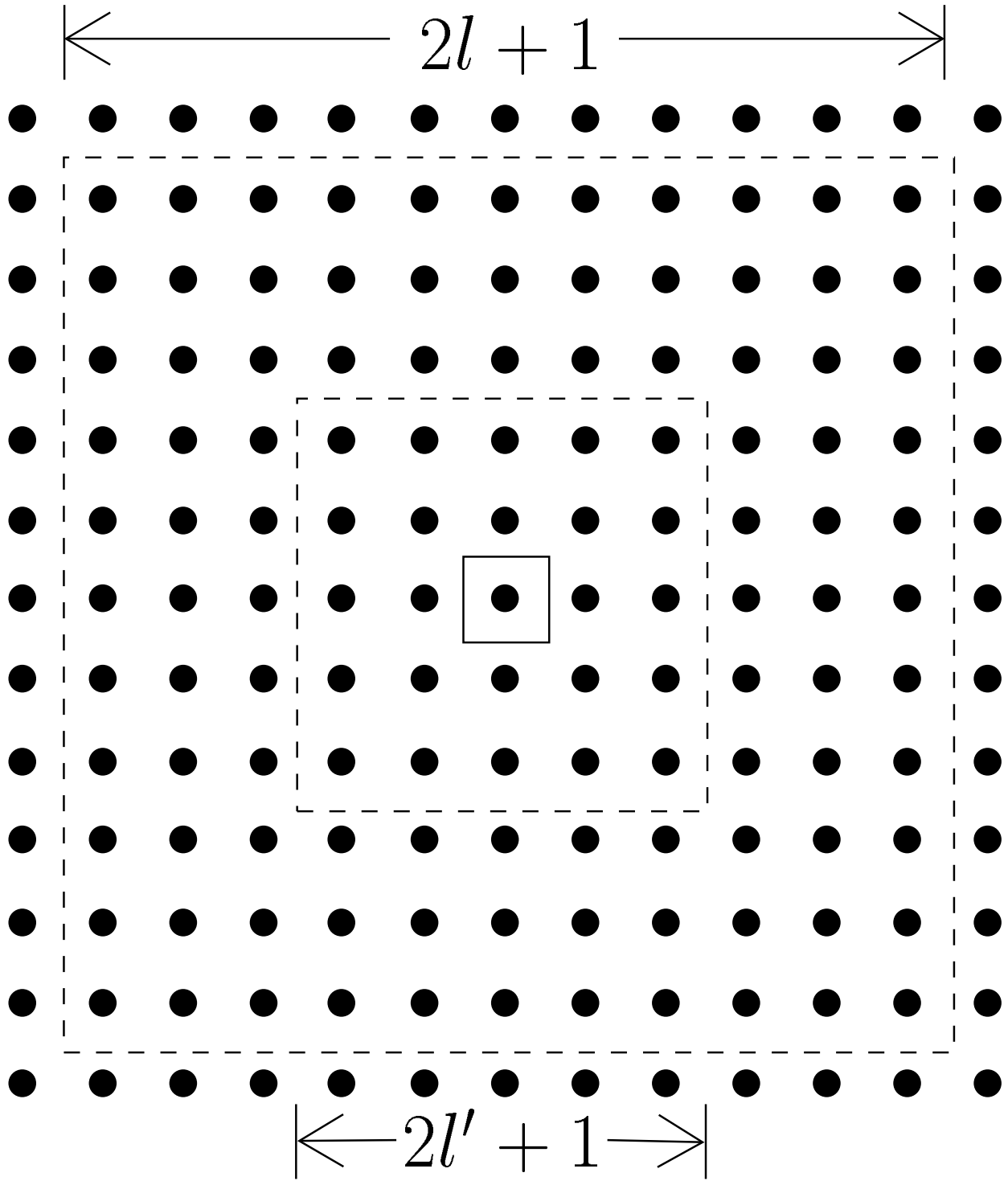}
\caption{Illustration of the local region $(l=5)$ with a central region $(l'=2)$.}
\label{fig:2}
\end{center}
\end{figure}

\clearpage

\begin{figure}
\begin{center}
\includegraphics[width=1.0\textwidth]{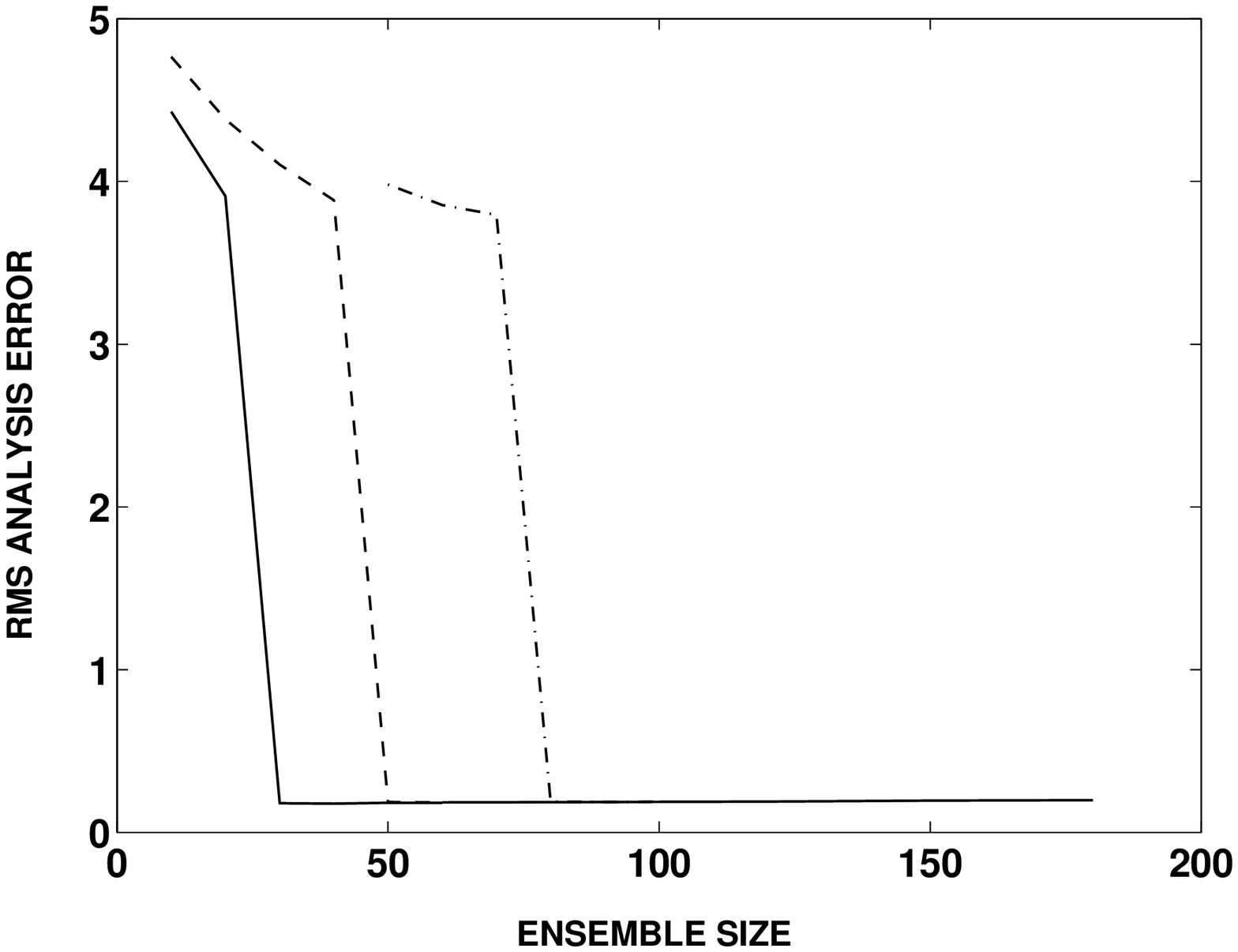}
\caption{The rms error of the full Kalman filter as function of the number of
ensemble members. Shown are the results for $M=40$ (solid line), $M=80$ (dashed
line), and $M=120$ (dotted-dashed line).} \label{fig:3}
\end{center}
\end{figure}

\clearpage

\begin{figure}
\begin{center}
\includegraphics[width=1.0\textwidth]{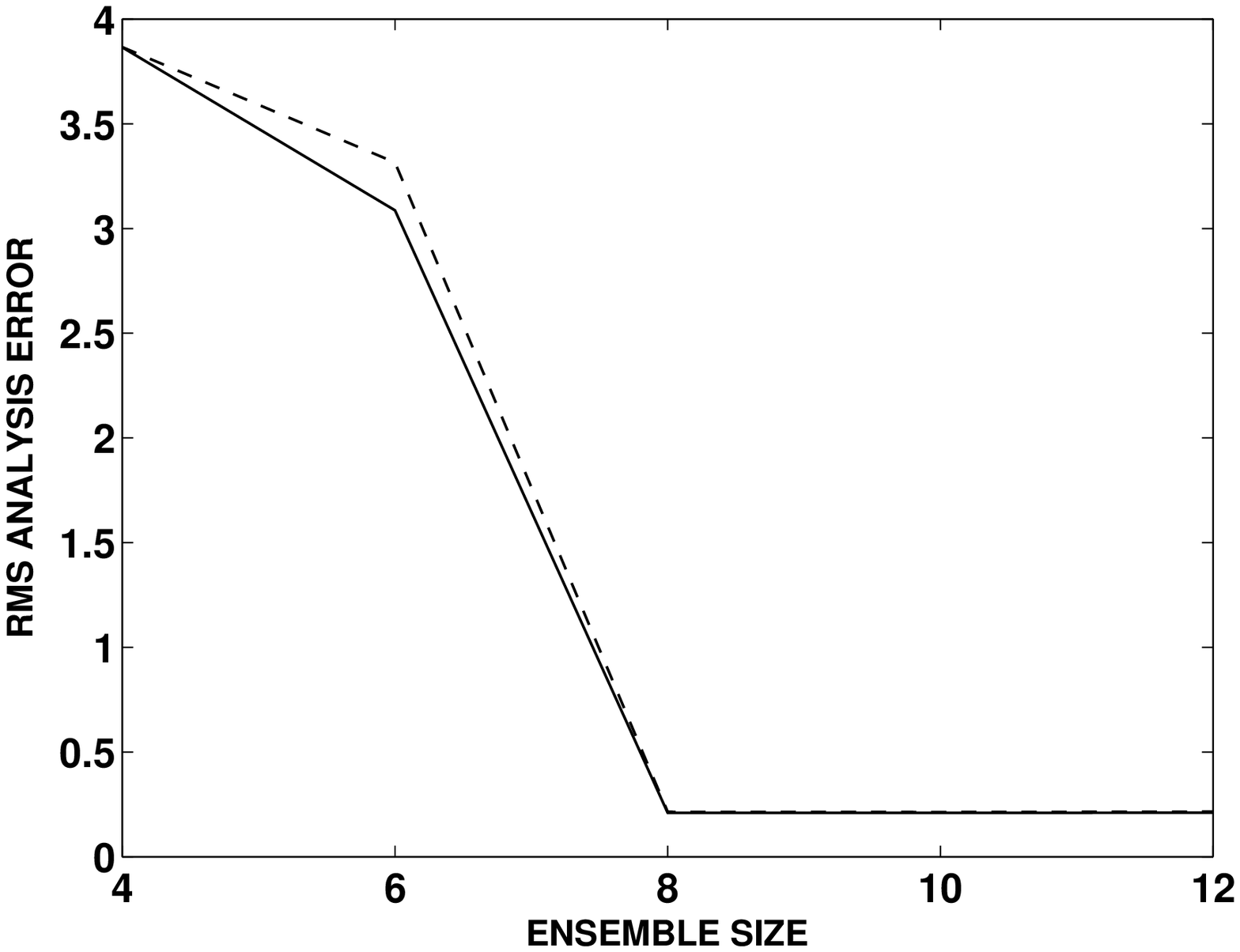}
\caption{The rms error of the local ensemble Kalman filter as function of the number
of ensemble members. Shown are the results for $M=40$ (solid line), $M=80$ (dashed
line), and $M=120$ (dotted-dashed line).} \label{fig:4}
\end{center}
\end{figure}

\clearpage

\begin{figure}
\begin{center}
\includegraphics[width=1.0\textwidth]{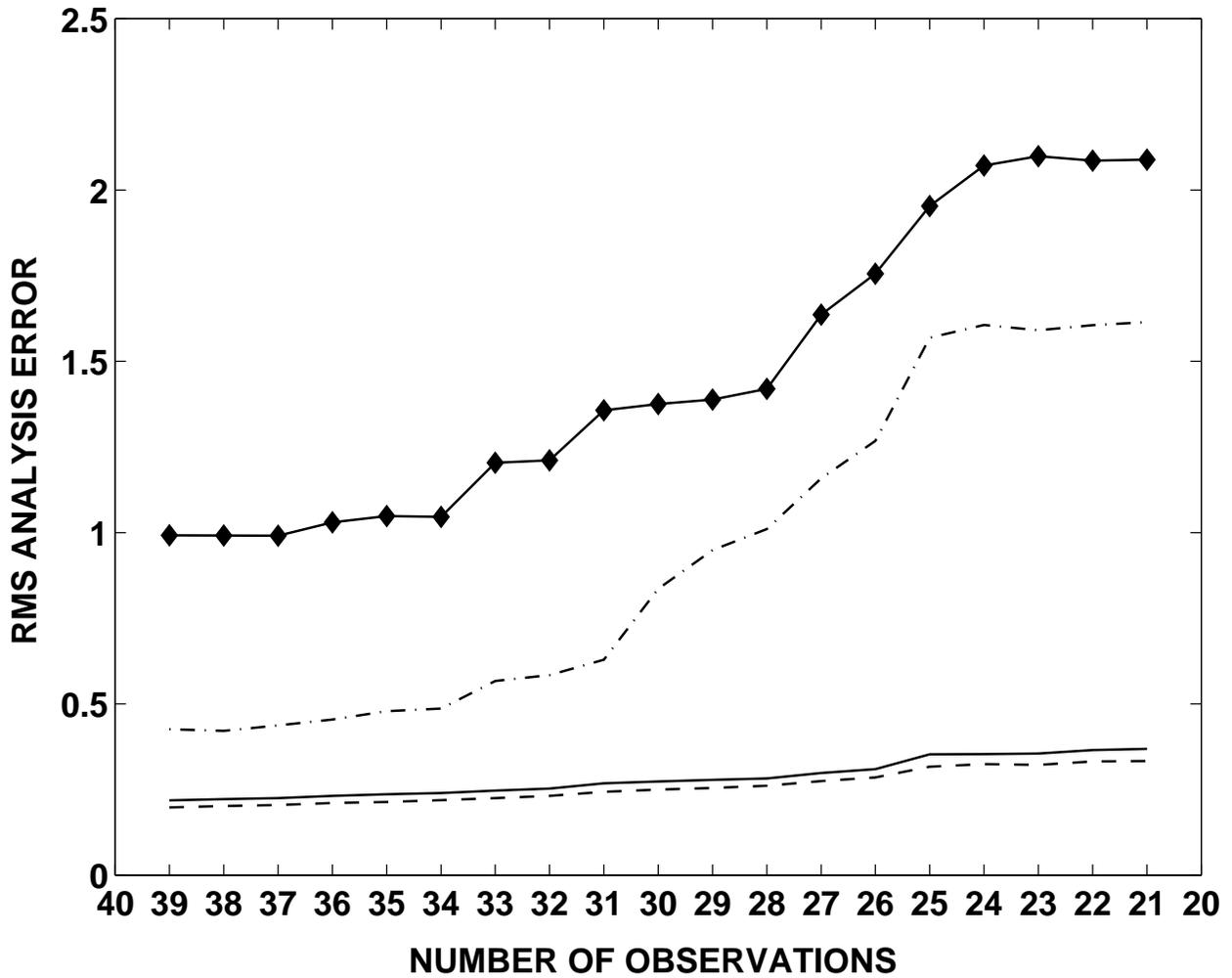}
\caption{The rms error of the different analysis schemes as function of the number
of observations. Shown are the results for the full Kalman filter [4\% variance
inflation] (dashed line), conventional scheme (dashed-dotted line), direct insertion
(solid line with diamonds), and the local ensemble Kalman filter [3\% variance
inflation] (solid line).} \label{fig:5}
\end{center}
\end{figure}

\clearpage

\begin{figure}
\begin{center}
\includegraphics[width=1.0\textwidth]{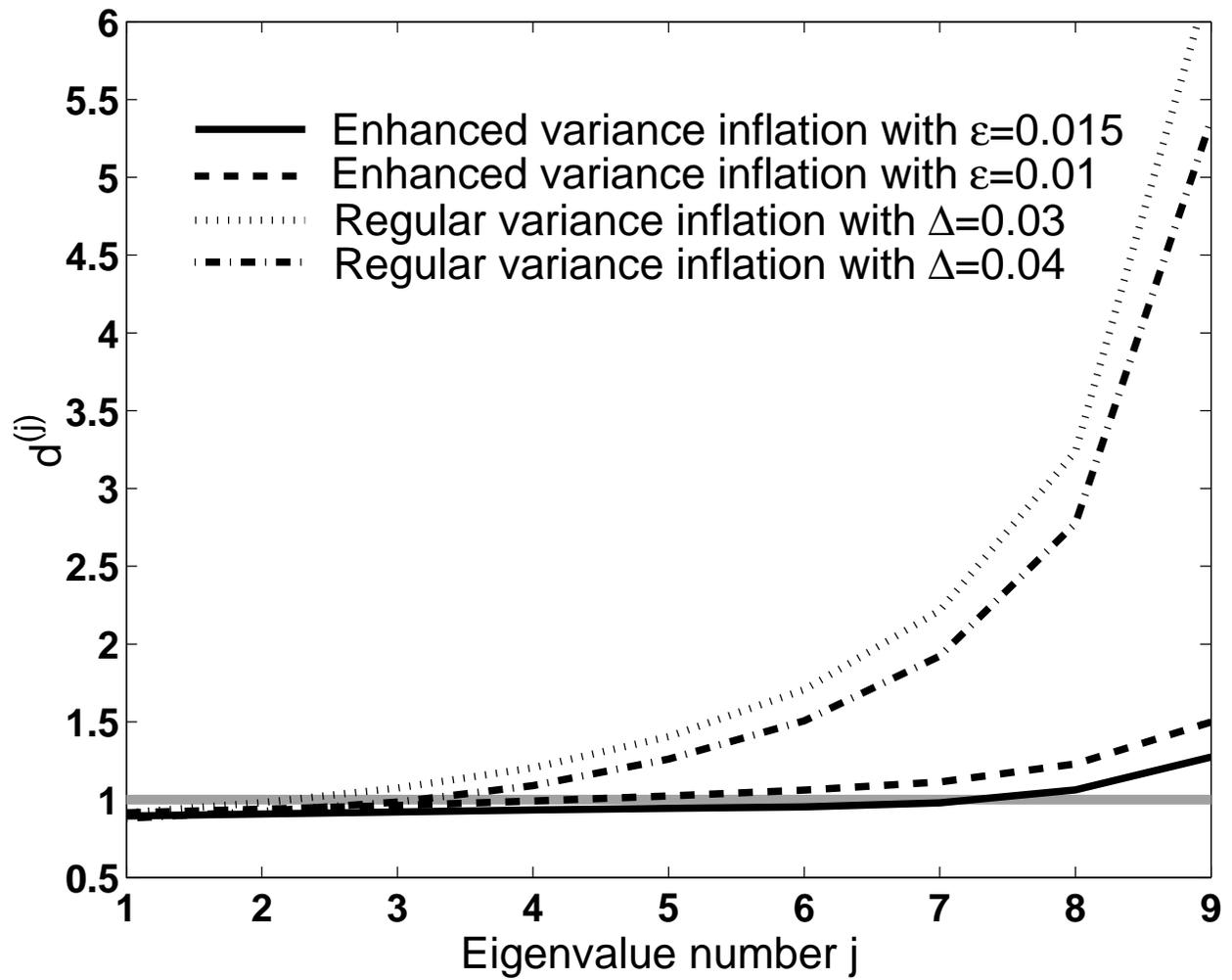}
\caption{The ratio $d^{(j)}$ at $m=1$ as function of $j$ for two different
values of $\varepsilon$ and $\Delta$.}
\label{fig:6}
\end{center}
\end{figure}

\clearpage

\begin{figure}
\begin{center}
\includegraphics[width=1.0\textwidth]{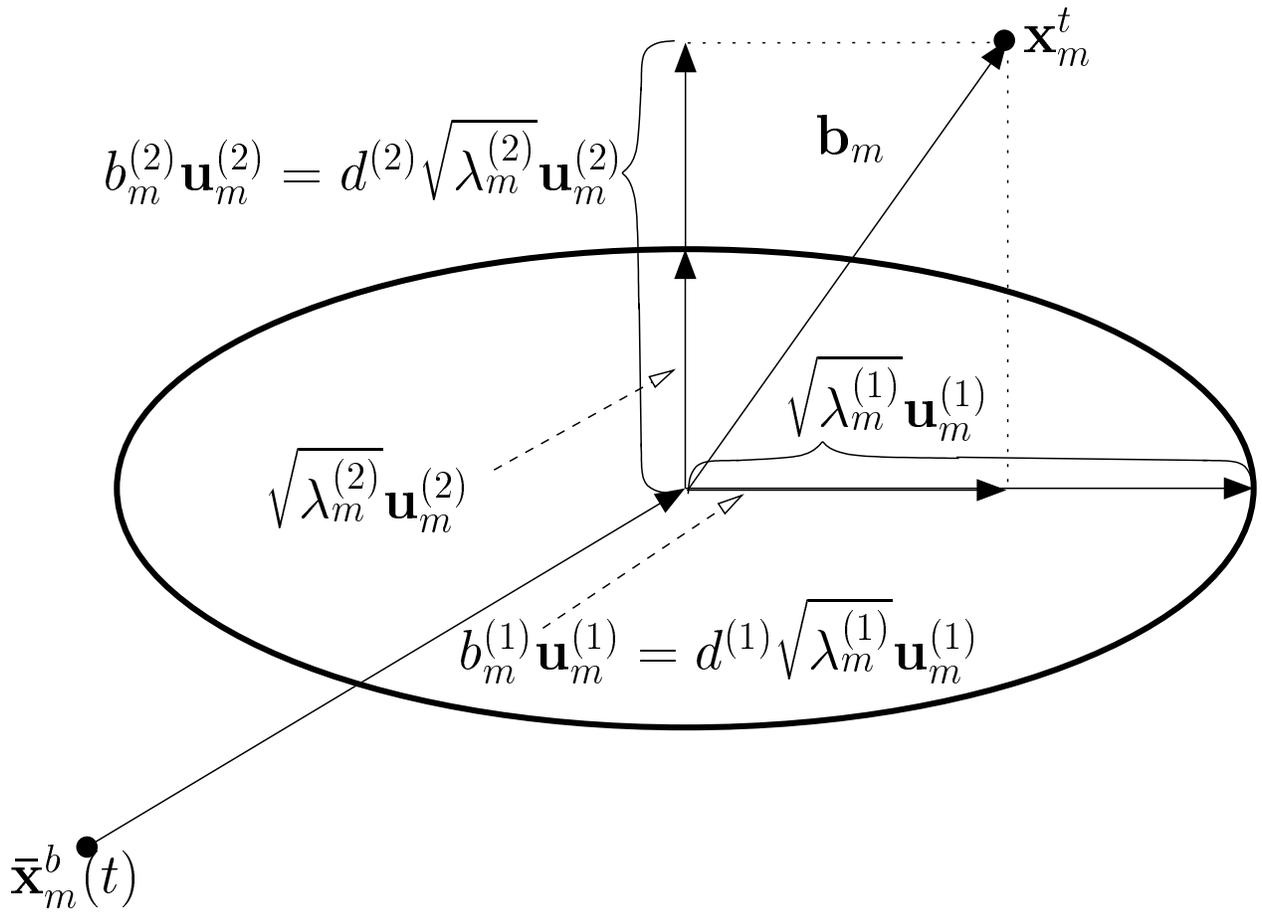}
\caption{Projection of the true background error, $\mathbf{b}_m$ on the
main axes of the probability ellipsoid.}
\label{fig:7}
\end{center}
\end{figure}

\clearpage

\begin{figure}
\begin{center}
\includegraphics[width=1.0\textwidth]{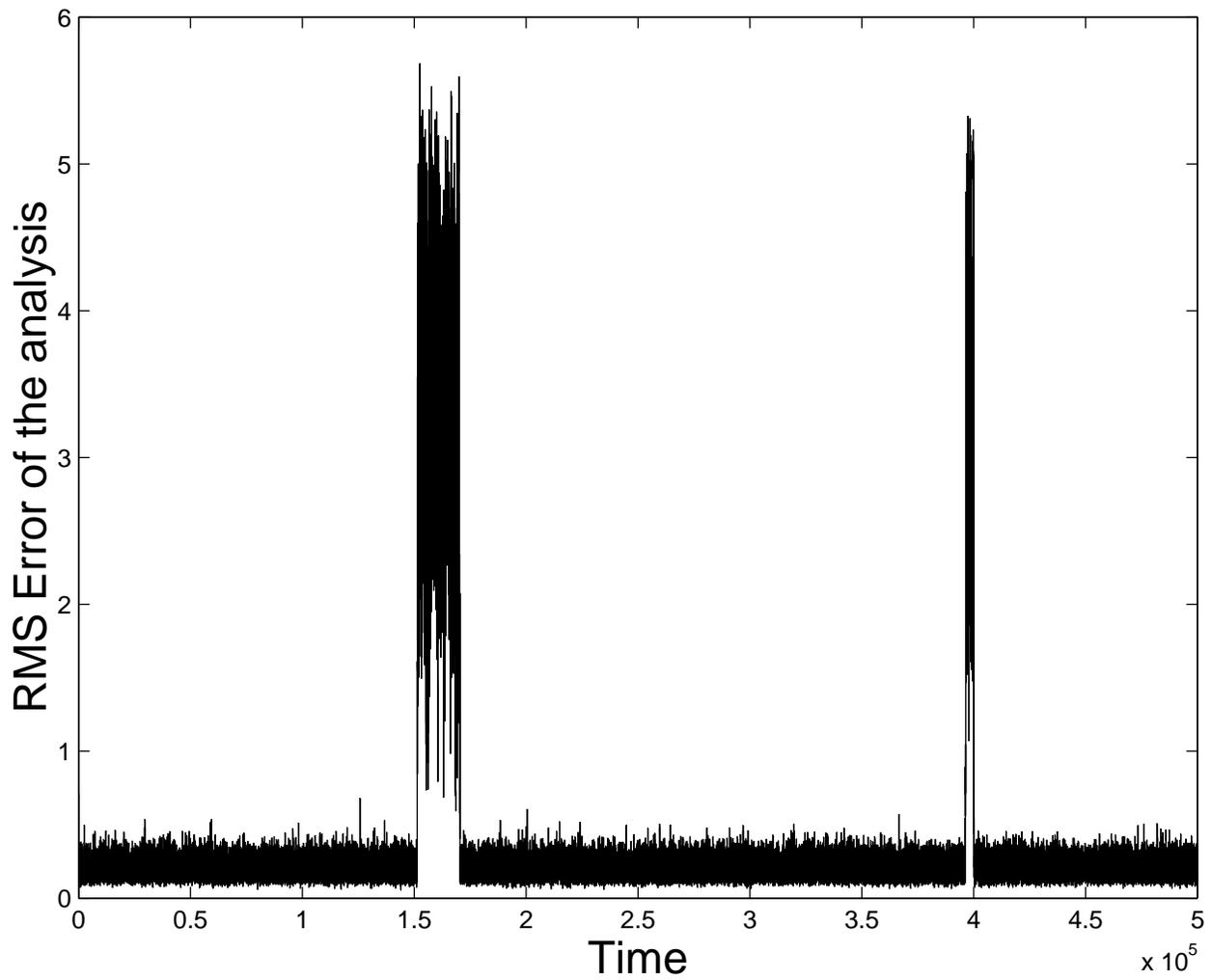}
\caption{The long time evolution of the rms analysis error for a set of
parameters that allow spikes to occur.
}
\label{fig:8}
\end{center}
\end{figure}

\clearpage

\begin{figure}
\begin{center}
\includegraphics[width=1.0\textwidth]{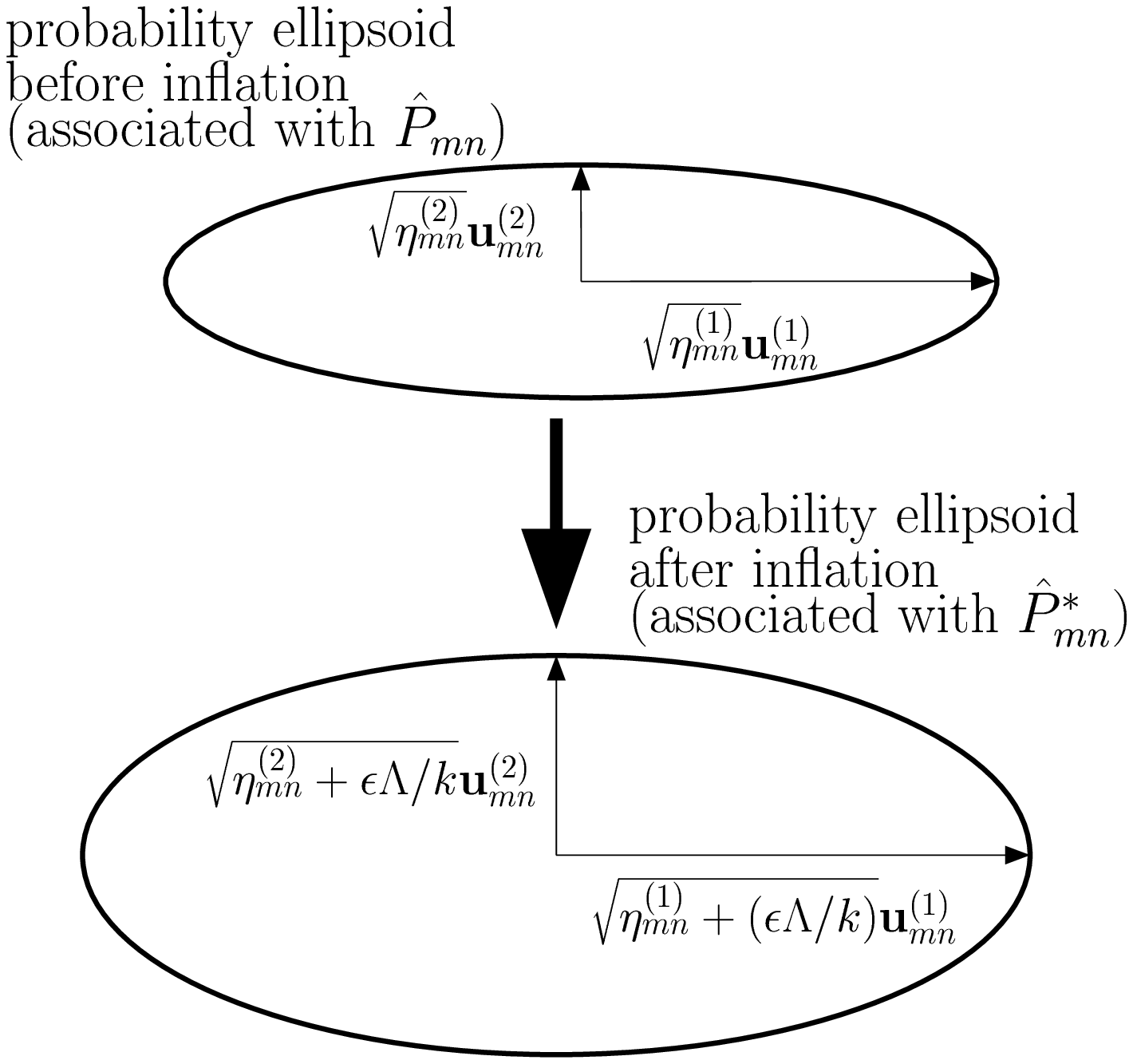}
\caption{The effect of the enhanced variance inflation (equation
\ref{eq:a3}) on the probability ellipsoid.  For the special case
$\hat{\mathbf{P}}_{mn}=\hat{\mathbf{P}}_{mn}^b$, $\eta_{mn}^{(1)}=
\lambda_{mn}^{(1)}$ and $\eta_{mn}^{(2)}=\lambda_{mn}^{(2)}$.}
\label{fig:9}
\end{center}
\end{figure}

\clearpage

\begin{figure}
\begin{center}
\includegraphics[width=1.0\textwidth]{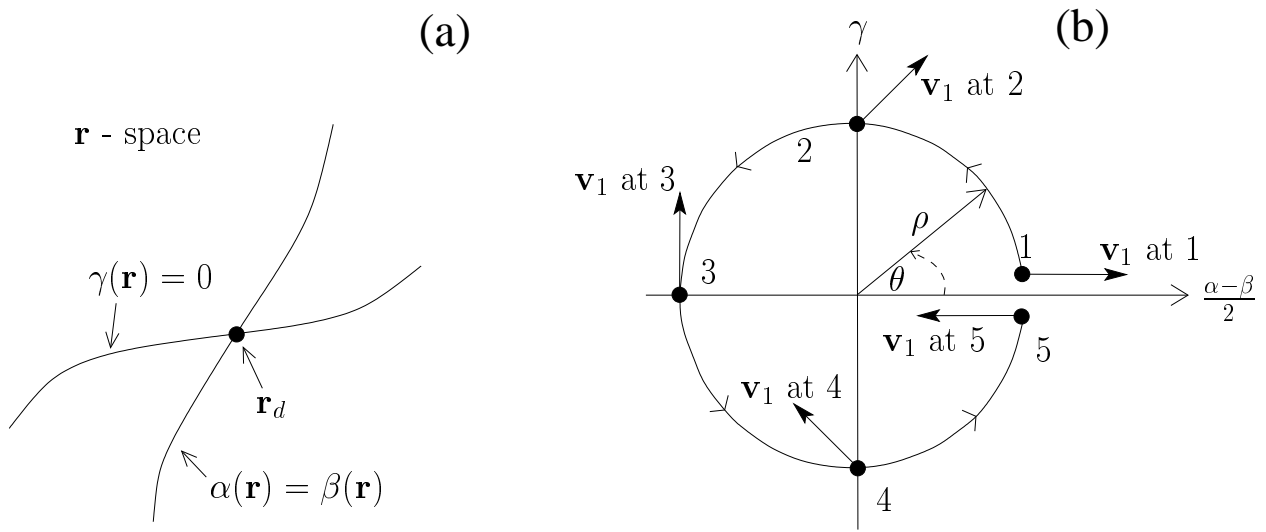}
\caption{(a) The curves $\gamma(\mathbf{r})=0$ and
$\alpha(\mathbf{r})=\beta(\mathbf{r})$ cross at a diabolical point
$\mathbf{r}_d$. (b) The eigenvector $\mathbf{v}_1$ flips by $180^0$ on one
circuit around the diabolical point.}
\label{fig:10}
\end{center}
\end{figure}

\end{document}